\PassOptionsToPackage{table,xcdraw,dvipsnames}{xcolor}

\documentclass[sigconf]{acmart}
\usepackage{booktabs}
\usepackage{arydshln}
\usepackage{amsmath}
\usepackage{multirow}
\usepackage[font=small,labelfont=bf]{caption}
\usepackage{array}

\newcolumntype{P}[1]{>{\raggedright\arraybackslash}p{#1}}
\newcommand{\set}[1]{\mathcal{#1}}
\newcommand{\mlm}{\texttt{MLM}}
\newcommand{\simt}{\texttt{SIM}}
\newcommand{\simtcls}{\texttt{SIM\textsubscript{CLS}}}
\newcommand{\simtmean}{\texttt{SIM\textsubscript{MEAN}}}
\newcommand{\nsp}{\texttt{NSP}}

\newcommand{\templateOne}{TP-NoTitle}
\newcommand{\templateTwo}{TP-Title}
\newcommand{\templateThree}{TP-TitleGenre}

\AtBeginDocument{%
  \providecommand\BibTeX{{%
    \normalfont B\kern-0.5em{\scshape i\kern-0.25em b}\kern-0.8em\TeX}}}

\copyrightyear{2020}
\acmYear{2020}
\setcopyright{acmcopyright}\acmConference[RecSys '20]{Fourteenth ACM Conference on Recommender Systems}{September 22--26, 2020}{Virtual Event, Brazil}
\acmBooktitle{Fourteenth ACM Conference on Recommender Systems (RecSys '20), September 22--26, 2020, Virtual Event, Brazil}
\acmPrice{15.00}
\acmDOI{10.1145/3383313.3412249}
\acmISBN{978-1-4503-7583-2/20/09}

\begin{document}

\title[What does BERT know about books, movies and music?]{What does BERT know about books, movies and music? Probing BERT for Conversational Recommendation}


\author{Gustavo Penha}
\affiliation{%
\institution{Delft University of Technology}
\city{Delft}
\country{Netherlands}}
\email{g.penha-1@tudelft.nl}

\author{Claudia Hauff}
\affiliation{%
\institution{Delft University of Technology}
\city{Delft}
\country{Netherlands}}
\email{c.hauff@tudelft.nl}


\begin{abstract}
Heavily pre-trained transformer models such as BERT have recently shown to be remarkably powerful at language modelling, achieving impressive results on numerous downstream tasks. It has also been shown that they implicitly store factual knowledge in their parameters after pre-training. Understanding what the pre-training procedure of LMs actually learns is a crucial step for using and improving them for \emph{Conversational Recommender Systems} (CRS). We first study how much off-the-shelf pre-trained BERT ``knows'' about recommendation items such as books, movies and music. In order to analyze the knowledge stored in BERT's parameters, we use different \emph{probes} (i.e., tasks to examine a trained model regarding certain properties) that require different types of knowledge to solve, namely \emph{content-based} and \emph{collaborative-based}. Content-based knowledge is knowledge that requires the model to match the titles of items with their content information, such as textual descriptions and genres. In contrast, collaborative-based knowledge requires the model to match items with similar ones, according to community interactions such as ratings. We resort to BERT's Masked Language Modelling (MLM) head to probe its knowledge about the genre of items, with cloze style prompts. In addition, we employ BERT's Next Sentence Prediction (NSP) head and representations' similarity (SIM) to compare relevant and non-relevant search and recommendation query-document inputs to explore whether BERT can, without any fine-tuning, rank relevant items first. Finally, we study how BERT performs in a conversational recommendation downstream task. To this end, we fine-tune BERT to act as a retrieval-based CRS. Overall, our experiments show that: \textit{(i)} BERT has knowledge stored in its parameters about the content of books, movies and music; \textit{(ii)} it has more content-based knowledge than collaborative-based knowledge; and \textit{(iii)} fails on conversational recommendation when faced with adversarial data.
\end{abstract}

\ccsdesc[300]{Information systems~Personalization}
\ccsdesc[500]{Information systems~Language models}
\ccsdesc[300]{Information systems~Recommender systems}

%
\keywords{conversational recommendation, conversational search, probing}

\maketitle


\section{Introduction}


Conversational Recommender Systems (CSR) have the potential to improve traditional recommender systems. For instance, they can provide better elicitation of a user's preferences, provide explanations of its suggestions, and also process a user's feedback related to its recommendations~\cite{jannach2020survey}. Given recent advances in machine learning techniques and the widespread deployment of voice-based agents such as Google Assistant and Amazon Alexa, there has been an uptake in conversational search \& recommendation research in the direction of natural language based approaches, as opposed to previous form-based and slot-filling techniques.

One important breakthrough in Natural Language Processing (NLP) is the use of heavily pre-trained transformers for language modelling, such as BERT~\cite{devlin2018bert} 
or T5~\cite{raffel2019exploring}. These pre-trained Language Models (LMs) are extremely powerful for many downstream tasks in NLP as well as IR, Recommender Systems (RecSys), Dialogue Systems (DS) and other fields---and have thus become an essential part of our machine learning pipelines. One advantage of these models is their capability to perform well on specific tasks and domains (that were \emph{not} part of their training regime) via fine-tuning, i.e. the retraining of a pre-trained model with just a few thousand labelled task- and/or domain-specific examples. Besides the power of such models to model human language, they have also been shown to store factual knowledge in their parameters~\cite{petroni2019language,roberts2020much}. For instance, we can extract the fact that the famous Dutch painter Rembrandt died in Amsterdam by feeding the prompt sentence \textit{"Rembrandt died in the city of \_\_\_\_"} to a pre-trained LM\footnote{This specific example works with both \texttt{bert-large-cased} and \texttt{roberta-large} in the fill-mask pipeline from the transformers library \url{https://huggingface.co/transformers/pretrained_models.html}.}, and use the token with the highest prediction score as the chosen answer.

Given the prevalence of such heavily pre-trained LMs for transfer learning in NLP tasks~\cite{tenney2019bert,qiu2020pre}, it is important to understand what the pre-training objectives are able to learn, and also what they fail to learn. Understanding the representations learned by such models has been an active research field, where the goal is to try and understand what aspects of language such models capture. Examples include analyzing the attention heads~\cite{clark2019does,michel2019sixteen}, or using probing tasks~\cite{jawahar2019does,tenney2019bert} that show which linguistic information is encoded. Such LMs have been successfully applied to different IR tasks~\cite{yang2019simple,yang2019end,sakata2019faq,Qu:2019:BHA:3331184.3331341}, but it is still unknown what exactly makes them so powerful in IR~\cite{camara2020diagnosing}. Unlike previous studies, we diagnose LMs here from the perspective of CRS. We focus on BERT~\cite{devlin2018bert} as its publicly released pre-trained models have been shown to be effective in a wide variety of NLP and IR tasks.


\begin{table*}[tbp]
\footnotesize
\caption{Input and output examples for the probing and downstream tasks considered in the movie domain. For the first task, \textbf{recommendation}, the user input is the history of seen movies, and the output is the recommendation for what to watch next. This task requires a model to match movies that are often seen together by different users---and thus are similar in a collaborative sense. We refer to this as collaborative-based knowledge. The second task, \textbf{search}, requires that a model matches descriptions of the item (item review) with the title. Similarly, the \textbf{genre} requires the model to match the genres of the items with their titles. We refer to this type of knowledge described in the second column as content-based. In \textbf{conversational recommendation} (the downstream task we focus on here), we see that knowing that "\textit{Pulp Fiction}" is a movie often seen by people who saw "\textit{Power Rangers}" (\textcolor{blue}{recommendation probe}), that it has a good soundtrack (\textcolor{red}{search probe}), and that it is from the genres "\textit{drama}" and "\textit{thriller}" (\textcolor{ForestGreen}{genre probe}) are helpful information to give a credible and accurate response.}

\label{table:introduction_table}

\begin{tabular}{@{}p{2cm}p{3cm}p{3cm}p{5cm}@{}}
\toprule
 & \textbf{Recommendation} & \textbf{Search and Genre} & \textbf{Conversational Recommendation}\\ \midrule
\textbf{User input} & Critters (1986) → NeverEnding Story, The (1984) → \textcolor{blue}{Power Rangers (1995)} → \textcolor{blue}{Turbo: A Power Rangers Movie (1997)}→ & \textbf{search} "[...] and there's the \textcolor{red}{music} in the movie: the \textcolor{red}{songs} Tarantino chose for his masterpiece fit their respective scenes so perfectly that most of those pieces of \textcolor{red}{music}." \textbf{genre} "\textcolor{ForestGreen}{drama, thriller}" &
"90's film with great \textcolor{red}{soundtrack}.[...] I thought \textcolor{blue}{Power Rangers in 1995} and then \textcolor{blue}{Turbo in 1997} were masterpieces of cinema, mind you {[}..{]}  I'm looking for movies from that era with great \textcolor{red}{music}. \textcolor{ForestGreen}{Dramas, thrillers}, road movies, adventure… Any genre (except too much romantic) will do ."\\ \hline
\textbf{System output} & Pulp Fiction (1994) & Pulp Fiction (1994) & You should see Pulp Fiction, Rock Star, [...]\\ \hline
\textbf{Task type} & probing & probing & downstream\\ \hline
\textbf{Knowledge}  & collaborative & content  & content and collaborative \\ 
\bottomrule
\end{tabular}

\end{table*}

Thus, our first research question (\textbf{RQ1}) is: \textit{How much knowledge do off-the-shelf BERT models store in their parameters about items to recommend?} We look specifically at movies, books and music due to their popularity, since many users frequently engage with recommenders in such domains\footnote{Some of the largest existent commercial recommender systems such as Netflix, Spotify and Amazon focus on such domains.}.

In order to provide a better intuition of our work, consider the examples in Table \ref{table:introduction_table}. Shown are examples (for the movie domain) of inputs and outputs for the different tasks considered in our work. In conversational recommendation, users engage in a conversation with the system to obtain recommendations that satisfy their current information need. This is the downstream task we are focused in this paper. The users often describe items that they have interacted with and enjoyed (\textcolor{blue}{{\textit{"Power Rangers in 1995 and then Turbo in 1997"}}}), and give textual descriptions of what they are looking for regarding the recommendation (\textcolor{red}{\textit{"film with great soundtrack"}} and \textcolor{ForestGreen}{\textit{"dramas, thrillers"}}). Such interactions can be categorized as having the intent of providing preferences~\cite{jannach2020survey}. We consider the knowledge of which items are often consumed together to be \emph{collaborative-based knowledge}, and we examine models for this through a recommendation probing task: \textit{given an item, find similar ones} (according to the community interaction data such as ratings from ML25M~\cite{harper2015movielens}), e.g. users who like \textit{"Power Rangers"} also like \textit{"Pulp Fiction"}. We consider the descriptions about the content of the items to be \emph{content-based knowledge}, and we examine models for this using a search probing task for which a review of the item has to be matched with the title of the item, and a genre probing task for which the genres of the movie have to be matched with the movie title.

To answer \textbf{RQ1}, we probe BERT models on \emph{content-based knowledge}, by using the predictions of BERT's Masked Language Modelling (MLM) head. We use knowledge sources to extract the information of the genre of the items, and generate prompt sentences such as \textit{"Pulp Fiction is a movie of the \_\_\_\_ genre."} similar to prior works~\cite{petroni2019language}, for which the tokens \textit{drama, thriller} should have high prediction scores in case the BERT model stores this information. In order to probe BERT models for the search and recommendation probing tasks, we introduce two techniques that do not require fine-tuning, and are able to estimate the match between two sentences. One technique is based on BERT's Sentence Representation Similarity (SIM), with the other based on BERT's Next Sentence Prediction (NSP) head. We generate the relevant recommendation prompt sentences with items that are frequently consumed together, and use both techniques to compare them against the non-relevant ones with items that are rarely consumed together. For example, the prompt \textit{“If you liked Pulp Fiction [SEP] you will also like Reservoir Dogs“}\footnote{Note that the \texttt{[SEP]} token is used by BERT as sentence separator, and we therefore use the next sentence predictor head as a next \emph{subsentence} predictor head.} should have a higher next sentence prediction score than the input \textit{“If you liked Pulp Fiction [SEP] you will also like To All the Boys I've Loved Before”}, since the first two movies co-occur more often than the second pair based on ratings data such as MovieLens~\cite{harper2015movielens}. For the search prompt, we generate relevant sentences by matching the title of the items with their respective reviews, a common approach to simulate product search~\cite{guo2019attentive,zamani2020learning}.

Our experimental results for \textbf{RQ1} reveal the following:
\begin{itemize}
    \item BERT has both collaborative-based and content-based knowledge stored in its parameters; correct genres are within the top-5 predicted tokens in 30\% to 50\% of the cases depending on the domain; reviews are matched to correct items 80\% of the times in the book domain when having two candidates; correct recommendation sentences are selected around 60\% of the time when having two candidates.
    \item BERT is more effective at storing content-based knowledge than collaborative-based knowledge.
    \item The NSP is an important pre-training objective for the search and recommendation probing tasks, improving the effectiveness over not using the NSP head by up to 58\%.
    \item BERT's effectiveness for search and recommendation probes drops considerably when increasing the number of candidates in the probes, especially for collaborative-based knowledge (i.e., a 35\% decrease in recall at the first position).
\end{itemize}    
Based on these findings, we next study how to use BERT for conversational recommendation and, more importantly, manners to infuse collaborative-based knowledge and content-based knowledge into BERT models as a step towards better CRS. We hypothesize that a model which is able perform well at search and recommendation probing tasks to be better for conversational recommendation. And thus, our second research question (\textbf{RQ2}) is: \textit{What is an effective manner to infuse additional knowledge for conversational recommendation into BERT?} Our experimental results show the following.

\begin{itemize}
    \item Our fine-tuned BERT is highly effective in distinguishing relevant responses and nonrelevant responses, yielding 36\% nDCG@10 improvement when compared to a competitive baseline for the downstream task.

    \item When faced with adversarially generated negative candidates that recommend random items, BERT's effectiveness degrades significantly (from 0.81 nDCG@10 to 0.06). 

    \item Infusing content-based and collaborative-based knowledge via multi-task learning during the fine-tuning procedure improves conversational recommendation.
\end{itemize}



\section{Related Work}

\subsubsection*{Conversational Search and Recommendation}
With the emergence of conversational agents, such as the Google Assistant and Apple's Siri---along with recent progress in machine learning for NLP, such as transfer learning with transformer models--- conversational search and recommendation has received an increased research interest. This new paradigm---where users engage in a natural language conversation with their chosen system to solve their information needs---yields the potential to solve and improve upon different aspects of search and recommendation. This can be achieved through clarification, elicitation, and refinement of a user's information need and providing recommendations, explanations and answers~\cite{jannach2020survey}. Despite recent work to conceptualize conversational search and recommendation~\cite{radlinski2017theoretical,trippas2018informing,vakulenko2018qrfa,azzopardi2018conceptualizing,radlinksi2019coached} and to improve techniques for related tasks~\cite{penha2020curriculum,zhang2018towards,li2018towards,rosset2020leading}, developing a full-blown system is still a major and open challenge.

As pointed out by~\citet{jannach2020survey}, the major actions a CRS must handle are: \textit{request}, where the system elicits the user's preferences; \textit{recommend}, where the system recommends items to the user, \textit{explain}, where the system provides explanations for the recommendations; and \textit{respond}, where the system answers questions that do not fall under the other actions. In this work, we focus on examining the potential of using pre-trained language models, specifically BERT~\cite{devlin2018bert}, for conversational recommendation---and where it excels or requires improvement. 

\subsubsection*{Pre-trained Language Models: Probing and Knowledge Infusion}

The extensive success of pre-trained transformer-based language models such as BERT~\cite{devlin2018bert}, RoBERTa~\cite{liu2019roberta}\footnote{RoBERTa is similar to BERT but its model is trained for longer on more data and without the NSP pre-training task.}, and T5~\cite{raffel2019exploring} can be attributed to the transformers' computational efficiency, the amount of pre-training data, the large amount of computations used to train such models\footnote{For instance, the RoBERTa model~\cite{liu2019roberta} was trained on 160GB of text using 1,024 32GB \textit{NVIDIA V100} GPUs} and the ease of adapting them to downstream tasks via fine-tuning. Given the remarkable success of such LMs, pioneered by BERT, researchers have focused on understanding what exactly such LMs learn during pre-training. For instance, by analyzing the attention heads~\cite{clark2019does,michel2019sixteen}, by using probing tasks~\cite{jawahar2019does,tenney2019bert} that examine BERT's representation to understand which linguistic information is encoded at which layer and by using diagnostic datasets~\cite{camara2020diagnosing}.

BERT and RoBERTa failed completely on 4 out of the 8 probing tasks that require reasoning skills in experiments conducted by~\citet{talmor2019olmpics}. The ``Always-Never'' probing task is an example of such a failure. Here, prompt sentences look like \textit{``rhinoceros [MASK] have fur''}, with candidate answers for this task being \textit{``never''} or \textit{``always''}.~\citet{petroni2019language} showed that BERT can be used as a competitive model for extracting factual knowledge, by feeding cloze-style prompts to the model and extracting predictions for its vocabulary.~\citet{jiang2019can} extended this work, demonstrating that using better prompt sentences through paraphrases and mined templates led to better extraction of knowledge from LMs.~\citet{roberts2020much} showed that off-the-shelf (i.e., pretrained LMs without fine-tuning) T5 outperformed competitive baselines for open-domain question answering.

Another line of work has focused on infusing different information in LM parameters to perform better at downstream tasks. One approach to do so is by having intermediary tasks before the fine-tuning on the downstream task~\cite{phang2018sentence}. The intuition here is that other tasks that are similar to the downstream task could improve the LM's effectiveness. It is still unknown why a combination of intermediate and downstream tasks is effective~\cite{pruksachatkun2020intermediate}. A similar approach is to continue the pre-training of the language model with domain-specific text corpora~\cite{gururangan2020don}.~\citet{wang2020meta} proposed a different approach inspired by multi-task learning~\cite{zhang2017survey} that grouped similar NLP tasks together. When infusing different types of knowledge into LMs, it is possible for some of the knowledge that was stored in its parameters to be erased, otherwise known as catastrophic forgetting~\cite{kirkpatrick2017overcoming}.~\citet{thompson2019overcoming} proposed a technique that regularizes the model when doing adaptation so that the weights are close to the pre-trained model.~\citet{wang2020k}~tackled this problem by proposing adapters, i.e., auxiliary neural modules that have different sets of weights, instead of sharing weights in a multi-task manner---and are effective when infusing different types of knowledge into LMs (such as factual and linguistic).

Instead of probing LMs for linguistic properties or general facts, we examine in our work LMs through the lens of conversational recommendation. Specifically, we look into recommendation, search and genre probes that require collaborative and content knowledge regarding items to be recommended. We then examine the effectiveness of the LMs for conversational recommendation---before and after infusing additional knowledge via multi-task learning.



\section{Method}

In this section, we briefly describe BERT, and then introduce in more detail our three types of probing tasks (genre, search and recommendation). We then turn towards our downstream task---conversational recommendation. 



\subsection{Background: BERT}

BERT ~\cite{devlin2018bert} is a transformer~\cite{vaswani2017attention} model that learns textual representations by conditioning on both left and right context for all layers. BERT was pre-trained for two different tasks, MLM and NSP. For MLM, 15\% of the tokens are replaced with a \texttt{[MASK]} token, and the model is trained to predict the masked tokens\footnote{More accurately, for the 15\% tokens, 80\% are replaced with \texttt{[MASK]}, 10\% of the time they are replaced with random tokens and the remaining 10\% the token is unchanged.}. For NSP, the model is trained to distinguish (binary classification) between pairs of sentences \texttt{A} and \texttt{B}, where 50\% of the time \texttt{B} is the next and 50\% it is not the next sentence (a random sentence is selected). The special token  \texttt{[CLS]} is added to every sentence during pre-training; it is used for classification tasks.  \texttt{[SEP]} is another special token that is used to separate sentence pairs that are packed together into a
single sequence. Additionally, there is a special learned embedding which indicates whether each token comes from sentence \texttt{A} or \texttt{B}. BERT was pre-trained using both English Wikipedia (2.5m words) and the BookCorpus~\cite{zhu2015aligning}, which contains the content of 11k books (800m words). Different pre-trained BERT models were released, changing the number of parameters (\textit{base} or \textit{large}) and tokenization (\textit{cased} or \textit{uncased}).


\begin{table*}[tbp]
\footnotesize
\caption{Examples of the probes used in this paper. We use off-the-shelf BERT's Masked Language Modelling (MLM) head for predicting tokens, BERT's Next Sentence Prediction (NSP) head for predicting if the underlined sentence is the most likely continuation of the sentence and BERT's last layer hidden representations (CLS pooled and MEAN pooled) for similarity between two texts (SIM). All probes require no fine-tuning, and thus indicate what BERT learns through its pre-training objectives. The knowledge source for recommendation prompts are interaction datasets, such as users' movie ratings. For search prompts, we use items' review data. No underline indicates query sentences, and \underline{underline} indicates document sentences. Relevant document sentences for a query have label 1 while non-relevant have label 0.}
\label{table:probe_examples}
\begin{tabular}{@{}lllll@{}}
\toprule
\textbf{Type} & \textbf{Prediction} & \textbf{Task} & \textbf{Prompt Examples} & \textbf{Labels}  \\ \midrule
\multirow{6}{*}{MLM} & \multirow{6}{*}{Token} & \multirow{6}{*}{Genre} & \templateOne: \textit{"It is a movie of the [MASK] genre."} & crime \\
 &  &  & \templateTwo :\textit{"Pulp Fiction is a [MASK] movie."} & crime  \\
 &  &  & \templateThree :\textit{"Pulp Fiction is a movie of the [MASK] genre ."} & crime \\ \cmidrule(l){4-5} 
 &  &  & \templateOne :\textit{"It is a book of the [MASK] genre."} & comic  \\
 &  &  & \templateTwo :\textit{"Palestine by Joe Sacco is a [MASK] book."} & comic \\
 &  &  & \templateThree :\textit{"Palestine by Joe Sacco is a book of the [MASK] genre."} & comic \\ \hline
\multirow{2}{*}{SIM} & \multirow{2}{*}{IsSimilar} & Recommendation & \textit{$\lbrace $("The Hobbit", \underline{"Lord of the Rings"}),} & \{1, 0\} \\
 &&& \textit{("The Hobbit", \underline{"Twilight"})}$\rbrace $ &  \\ \cmidrule(l){3-5} 
 &  & Search & \textit{$\lbrace $("The book is not about the murder {[}...{]}", "\underline{The Brothers Karamazov}"),} & \{1, 0\} \\
 &  & & \textit{("It gives a brilliant picture of three bright young people {[}...{]}", "\underline{The Brothers Karamazov.}") $\rbrace $} & \\\hline
\multirow{2}{*}{NSP} & \multirow{2}{*}{IsNext} & Recommendation & $\lbrace $ \textit{"If you liked The Hobbit, [SEP] \underline{you will also like Lord of the Rings}",} & \{1, 0\} \\
& & & \textit{"If You liked The Hobbit, [SEP] \underline{you will also like Twilight}"$\rbrace $} & \\ \cmidrule(l){3-5} 
&&Search&$\lbrace $ \textit{"The book is not about the murder  {[}...{]}  [SEP] \underline{The Brothers Karamazov.}}",&\{1, 0\}\\ 
&&&\textit{"It gives a brilliant picture of three bright young people {[}...{]} [SEP] \underline{The Brothers Karamazov.} "$\rbrace $} & \\ 
 \bottomrule
\end{tabular}
\end{table*}

\subsection{Genre Probes}
We resort to genre (i.e. a style or category of the item such as \textit{comedy}) probes to extract and quantify the knowledge stored in language models about the recommended items. Using knowledge sources that contain an item's title and its respective genres, e.g. "\textit{Los miserables by Victor Hugo}" $\longrightarrow$ \textit{"romance, fiction, history"}\footnote{We can extract this information from user generated tags to books for example.}, we create prompt sentences for each item with the genre as the masked token. Since we use the MLM head to predict masked tokens, we refer to this type of probing as \mlm{}. We use three manually defined prompt sentence templates (cf. Table~\ref{table:probe_examples}, first row, for examples of each template type) inspired by~\cite{jiang2019can} for the \mlm{} probe to investigate what BERT can do with different template sentences:
\begin{description}
    \item[\templateOne:] we do not provide the item to the probe, but only provide the domain of the item;
    \item[\templateTwo:] we use both the title of the item and its domain; \item[\templateThree:] we provide the item title, domain as well as an additional phrase \textit{"of the genre"} indicating that we are looking specifically for the genre of the item.
\end{description}
The underlying assumption of this probing technique is that if the correct tokens are ranked higher by the language model, it has this knowledge stored in its parameters about the item. We evaluate the amount of knowledge stored in the model by counting the number of correctly ranked labels as the most probable in the first and first 5 positions, i.e. R@1 and R@5. Since the template sentences are not exhaustive, our manually selected templates offer only a lower bound on the amount of knowledge stored in the language model. 

\subsection{Recommendation and Search Probes}
\label{section:rec_and_search_probes}

In order to probe a LM's capacity to rank relevant items in recommendation and search scenarios we now introduce two probing techniques (\simt{} and \nsp{}). Like the genre probe, these two techniques do not require any fine-tuning to quantify the LM's ranking effectiveness. We were inspired by methods to calculate the matching degree between two sentences, in a non-supervised way~\cite{zhang2019bertscore}. While \simt{} uses the representations directly to calculate the matching degree, \nsp{} relies on the fact that this pre-training BERT head was designed to understand the relationship between the two sentences, something not directly captured by the MLM training~\cite{devlin2018bert}.

Using both techniques, we compare prompt sentences (the template and prompt generation is explained shortly) that represent either a `query' or a `document'. The query sentences take input from the user side (for search this is the item description, and for recommendation this is the history of rated items), and the document sentences contain a possible answer from the system to this input (the item to be recommended). We refer to relevant document sentences as the ones that are relevant items for the query sentence. Non-relevant document sentences are randomly sampled.

\subsubsection{Probe Based on Similarity (SIM)}

\simt{} ranks document sentences for a query sentence based on the representations learned by the LM: we calculate the dot similarity between the query sentence and document sentences using either the CLS token representation (\simtcls{}), or the average pooling of all tokens (\simtmean{}). More formally:
\begin{equation}
  SIM_{CLS} = BERT_{CLS}(query\_sentence) \cdot BERT_{CLS}(document\_sentence)  
\end{equation}
where $BERT_{CLS}(s)$ means extracting the representation of the CLS token in the last layer, and 

\begin{eqnarray}
&& SIM_{MEAN} = BERT_{MEAN}(query\_sentence)\nonumber\\
&& \qquad\qquad\qquad \cdot BERT_{MEAN}(document\_sentence)
\end{eqnarray}

where $BERT_{MEAN}(s)$ means extracting the representations of each token in the last layer by taking the average.

\subsubsection{Probe Based on Next Sentence Prediction Head (NSP)}
\nsp{} ranks document sentences for a query sentence based on the likelihood of the document sentence being a next sentence for the query sentence. Stated formally: 
\begin{equation}
  NSP = BERT_{NSP}(query\_sentence \; | \; [SEP] \; | \; document\_sentence)
\end{equation}
where $|$ indicates the string concatenation operator. This probe technique also does not require any fine-tuning of BERT.

\subsubsection{Templates and Prompt Generation} \label{rec_and_search_templates}

Having defined our probing techniques, we now discuss how to generate the prompts for the recommendation and search probes, along with the templates we used. Based on the knowledge extracted from rating and review datasets, we create \emph{prompt sentences} in a similar manner to how previous work extracted knowledge from other data sources~\cite{petroni2019language,petroni2020context}. 

For the recommendation probe, the query sentence is built using an item that was rated by a user $u$, and the relevant document sentence is another item rated by $u$ as well. The non-relevant document sentences are items that were not rated by $u$, and are sampled based on  the item's popularity. Since we do not have access to negative feedback on items, we use a common assumption in the offline evaluation of recommender systems that a randomly sampled item is not relevant~\cite{10.1145/2043932.2043996}. The assumption for the recommendation \& search probes is that a model that has higher similarity between the query sentence and the relevant document sentence has knowledge regarding which items are consumed together. For instance, see the \simt{} recommendation example in Table \ref{table:probe_examples}---a successful collaborative-filtering recommender system would display a higher similarity between \textit{"The Hobbit"} and \textit{"Lord of the Rings"} (items extracted from the user ratings' history) than the similarity between \textit{"The Hobbit"} and \textit{"Twilight"} (an item not relevant to the given user). Conversely, for the \nsp{} probes, we expect the next sentence prediction from the relevant document sentence to be higher than the non-relevant ones. Using the same user as an example, the next sentence prediction score for the relevant query-document sentence \textit{"If you liked The Hobbit [SEP], you will also like Lord of the Rings"} should be higher than the non relevant query-document sentence \textit{"If You liked The Hobbit [SEP], you will also like Twilight"}.
    
For the search probe, the query sentence is built using entire reviews from the items, whereas the relevant document sentence is the title of the item for which the review was written\footnote{We remove the titles of the items from the reviews to make the task more challenging.}. Unlike the recommendation probe, we generate non-relevant combinations of query and documents by matching item titles (documents) with random reviews (queries) that are not for the item in question. We use review data to simulate product search inspired by previous works~\cite{zamani2020learning,ai2017learning,guo2019attentive,van2016learning}. For instance, we expect that the \simt{} and \nsp{} scores between the item \textit{"The Brothers Karamazov"} and one of its review texts \textit{"The book is not about the murder [...]"} to be higher than the scores between the item \textit{"The Brothers Karamazov"} a randomly sampled review.

\subsection{Infusing Knowledge into LMs for Conversational Recommendation}

Finally, we discuss our downstream task, i.e. the task we aim to solve better with knowledge gained from our probes. Let us first define how to use BERT as an end-to-end retrieval-based conversational recommender system by formally defining the problem, before discussing the infusion of knowledge into a pre-trained LM.

\subsubsection{Conversational Recommendation}
Given a historical dialogue corpus and a conversation (i.e. the user's current utterance and the conversation history), the downstream task of conversational recommendation with a retrieval-based system is defined as the ranking of the most relevant response available in the corpus~\cite{wu2017sequential,yang2018response,tao2019one,penha2020curriculum}. This setup relies on the fact that a large corpus of historical conversational data exists, and adequate replies (that are coherent, well-formulated, complete with item recommendations) to user utterances can be found in it. Formally, let $\set{D}=\{(\set{U}_i, \set{R}_i, \set{Y}_i)\}_{i=1}^{N}$ be a conversational recommendation data set, consisting of $N$ triplets: dialogue context, response candidates and response labels. The dialogue context $\set{U}_i$ is composed of the previous utterances $\{u^1, u^2, ... , u^{\tau}\}$ at the turn $\tau$ of the dialogue. The candidate responses $\set{R}_i = \{r^1, r^2, ..., r^k\}$ are either the true response ($u^{\tau+1}$) or negative sampled candidates\footnote{In a production setup the ranker would either retrieve responses from the entire corpus, re-rank the responses retrieved by a recall-oriented retrieval method, or re-rank responses given by a language generation module.}. The relevance labels $\set{Y}_i = \{y^1, y^2, ..., y^k\}$ indicate the responses' binary relevance scores: 1 if $r = u^{\tau+1}$, and 0 otherwise. The objective is then to learn a ranking function $f(.)$ that is able to generate a ranked list for the set of candidate responses $\set{R}_i$ based on their predicted relevance scores $f(\set{U}_i,r)$. To fine-tune BERT, we make predictions as follows: 

\begin{equation}
f(\set{U}_i,r) = FFN(BERT_{CLS}(u^1 \; | \; [SEP] \; | \; u^2 \; | \; ... \; | \; u^{\tau} \; | \; [SEP] \; | \; r)),
\end{equation}

where $|$ indicates the concatenation operation\footnote{An example of input sequence for BERT is: \textit{"Good morning, I am looking for a good classic today. [SEP] What type of movie are you looking for today? [SEP] I enjoyed Annie (1982) [SEP] okay no problem. If you enjoyed Annie then you will love You've Got Mail (1998)"}} and $FFN$ is a linear layer. We train it for binary classification using cross entropy as the loss function. 

\subsubsection{Infusing Knowledge into LMs}

In order to infuse content-based and collaborative-based knowledge into BERT, we resort to multi-task learning~\cite{zhang2017survey}. In addition to fine-tuning BERT only for the conversational data, we also consider interleaving batches of different tasks. $BERT_{rec}$ interleaves training instances of the downstream task, with the recommendation \nsp{} probing task and (and analogously $BERT_{search}$ interleaves the downstream task with search \nsp{}). Multi-task learning is challenging as the order of the tasks~\cite{pentina2015curriculum} and the weighting~\cite{Kendall_2018_CVPR} for each tasks have a large impact on model's quality; we leave such analyses as future work and resort to equal weights and interleaved batches.





\section{Experimental Setup}

\subsection{Data Sources}

We use English language data\footnote{The data we created for this work as well as all our code are publicly available at \url{https://github.com/Guzpenha/ConvRecProbingBERT}.} from three different domains in order to generate the templates for our probes: 
\begin{description}
\item[Books:] we use the publicly available GoodReads\footnote{\url{https://sites.google.com/eng.ucsd.edu/ucsdbookgraph/home}} ~\cite{wan2018item} dataset that contains over 200M interactions from the GoodReads community. We extract ratings, reviews and genres.

\item[Movies:] we use the publicly available ML25M\footnote{\url{https://grouplens.org/datasets/movielens/25m/}}~\cite{harper2015movielens} dataset that contains 25M interactions from the MovieLens community. We extract ratings and genres. Since ML25M does not have any review data, we crawled reviews for movies that were rated in ML25M from IMDB. We collected a maximum of 20 reviews for each movie from the ML25M data. This resulted in a total of 860k reviews (av. length of 84.22 words) and an average of 13.87 reviews per movie. 
\item[Music:] We use the \textit{"CDs and Vinyl"} subset of the publicly available Amazon reviews\footnote{\url{https://nijianmo.github.io/amazon/index.html}}~\cite{ni2019justifying} dataset which contains 2.3m interactions. We extract ratings, reviews and genres.
\end{description}

For all the probes in this paper (genre, search and recommendation) we generate 100k probing instances, with the exception of movies in the genre probing task for which we have access to only approximately 60k movies (the number of movies in the ML25M dataset). For the genre probing task, we have on average 3.6, 1.8 and 1.4 genres for the books, movies and music domains and a total of 16, 20 and 284 distinct genres respectively.


Inspired by previous work that uses online forums as a source of rich conversational data~\cite{penha2019introducing,qu2018analyzing}, 
we extract conversational recommendation data for the three domains from reddit forums: \textit{/r/booksuggestions}, \textit{/r/moviesuggestions} and \textit{/r/musicsuggestions}\footnote{\url{https://www.reddit.com/r/booksuggestions/}, \url{https://www.reddit.com/r/moviesuggestions/} and \url{https://www.reddit.com/r/musicsuggestions/}} on March 17, 2020. They include multi-turn conversations where an information-seeker is looking for recommendations, and an information-provider gives suggestions through natural language conversations.\footnote{See the conversational recommendation example from Table~\ref{table:introduction_table} which comes from this dataset.}

Additionally, we use the ReDial dataset~\cite{li2018towards} which was collected using crowd workers, and includes dialogues of users seeking and providing recommendations in the movies domain. We use this dataset due to the annotated movie identifiers that are mentioned in each utterance, which is not available for the Reddit data. This allows us to create adversarial examples (see Table~\ref{table:redial_adv_examples} for a concrete example) that require the model to reason about different items to be recommended, while the rest of the response remains the same. The statistics of the data used for conversational recommendation are shown in Table~\ref{table:crr_data_stats}. For the music domain, there is a limited number of conversations available (the \textit{musicsuggestions} subreddit has only 10k users, compared to the 292k users of the \textit{booksuggestions} subreddit). 

For dialogue datasets, we generate 50 candidate responses for every context by querying all available responses with BM25~\cite{10.1561/1500000019} using the context as a query. This setup is in line with prior works~\cite{yang2018response}.
 
 \begin{table}[h]
    \centering
    \small
     \caption{Statistics of the conversational recommendation datasets. We use dialogues extracted from three subreddits: \textit{/r/booksuggestions}; \textit{/r/moviesuggestions}; and \textit{/r/musicsuggestions}. We also experiment with ReDial~\cite{li2018towards} due to its exact matches with movies.}
     \label{table:crr_data_stats}
    \begin{tabular}{@{}lllll@{}}
        \toprule
         & Books & Movies & Music & ReDial\\ \midrule
        \# $\mathcal{U}$--$r$ pairs & 157k & 173k & 2k & 61k \\
        \# candidates per $\mathcal{U}$ & 50 & 50 & 50 & 50\\
        Avg \# turns & 1.11 & 1.08 & 1.11 &  3.54\\
        Avg \# words per $u$ & 103.37 & 124.93 & 74.17 &  71.11\\
        Avg \# words per $r$ & 40.10 & 23.39 & 38.84 & 12.58\\
        \bottomrule
        \end{tabular}
\end{table}

\subsection{Implementation Details}
 
We use the BERT and RoBERTa PyTorch transformers implementations\footnote{\url{https://github.com/huggingface/transformers}}. When fine-tuning BERT for conversational recommendation, we employ a balanced number of relevant and non-relevant context and response pairs. We resort to BERT's default hyperparameters, and use the \textit{large} cased models; we fine-tune them with the Adam optimizer~\cite{kingma2014adam} with a learning rate of $5e-6$ and $\epsilon = 1e-8$. We employ early stopping using the validation nDCG. For the conversational recommendation task, we also employ as baselines traditional IR methods: QL~\cite{10.1145/290941.291008}, and QL with RM3~\cite{10.1145/383952.383972}. We use the pyserini\footnote{\url{https://github.com/castorini/pyserini/}} implementation of QL and RM3, and use the context as query and candidate responses as candidate documents. In addition, we compare BERT against strong neural baselines for the task: DAM~\cite{zhou2018multi}\footnote{\url{https://github.com/baidu/Dialogue/tree/master/DAM}}, and MSN~\cite{yuan2019multi}\footnote{\url{https://github.com/chunyuanY/Dialogue}}, which are interaction-based methods that learn interactions between the utterances in the context and the response with attention and multi-hop selectors, respectively. We fine-tune the hyperparameters for the baseline models (QL, RM3, DAM, and MSN) using the validation set.
 
 \begin{table}
    \centering
    \small
    \caption{Results for $BERT$ genre MLM probe. Bold indicates statistical significant difference over all other sentence types using a paired t-test with confidence level of 0.95 and Bonferroni correction.}    
    \label{table:results_genre_probe}
    \begin{tabular}{@{}lrrrrrr@{}}    
    \toprule
     & \multicolumn{6}{c}{Genre probes} \\ \cmidrule(lr{1em}){2-7}
     & \multicolumn{2}{c}{Books} & \multicolumn{2}{c}{Movies} & \multicolumn{2}{c}{Music} \\ \cmidrule(lr{1em}){2-7}
    Template & \multicolumn{1}{l}{$R@1$} & \multicolumn{1}{l}{$R@5$} & \multicolumn{1}{l}{$R@1$} & \multicolumn{1}{l}{$R@5$} & \multicolumn{1}{l}{$R@1$} & \multicolumn{1}{l}{$R@5$} \\ \midrule
    \templateOne & 0.067 & 0.259 & 0.067 & 0.395 & 0.074 & 0.412 \\ 
    \templateTwo & 0.031 & 0.119 & 0.058 & 0.258 & 0.139 & 0.346 \\ \hline
    \templateThree & \textbf{0.109} & \textbf{0.296} & \textbf{0.179} & \textbf{0.505} & \textbf{0.156} & 0.412 \\ \bottomrule
    \end{tabular}
 \end{table}

\begin{table*}[tbp]
\caption{Examples of $BERT$ predictions for each of the domains when probing it with the MLM head for item genres. BERT is able to match domains with common genres (\templateOne{} template), e.g. books with fantasy and music with rock. Prompt sentences that indicates to BERT it is looking for the genre of items (\templateThree{} as opposed to \templateTwo{}) yields better predictions as they avoid general descriptions, e.g. "television, 2003, japanese".}
\label{table:bert_cat_examples}
\footnotesize
\begin{tabular}{@{}lllll@{}}
\toprule
 & \textbf{Sentence} & \multicolumn{1}{c}{\textbf{Genre Prompt}} & \multicolumn{1}{c}{\textbf{Answer}} & \multicolumn{1}{c}{\textbf{Predicted (top 2)}} \\ \midrule
\multirow{3}{*}{\rotatebox[origin=c]{90}{Books}} & \templateOne & It is a book of the genre \_\_\_\_\_. & fiction fantasy & \textbf{fantasy} {[}0.183{]}, romance {[}0.133{]} \\
 & \templateTwo & The Wind-Up Bird Chronicle by {[}...{]} is a \_\_\_\_\_ book. & fiction fantasy & comic {[}0.072{]}, japanese {[}0.046{]} \\
 & \templateThree & The Wind-Up Bird Chronicle by {[}...{]} is a book of the genre \_\_\_\_\_. & fiction fantasy & \textbf{fantasy} {[}0.600{]}, horror {[}0.048{]} \\ \hline
\multirow{3}{*}{\rotatebox[origin=c]{90}{Movies}} & \templateOne & It is a movie of the genre \_\_\_\_\_. & action adventure & horror {[}0.083{]}, \textbf{action} {[}0.058{]} \\
 & \templateTwo & I, Robot (2004) is a \_\_\_\_\_ movie. & action adventure & tv {[}0.164{]}, television {[}0.162{]} \\
 & \templateThree & I, Robot (2004) is a movie of the genre \_\_\_\_\_. & action adventure & robot {[}0.548{]}, horror {[}0.086{]} \\ \hline
\multirow{3}{*}{\rotatebox[origin=c]{90}{Music}} & \templateOne & It is a music album of the genre \_\_\_\_\_. & rock & pop {[}0.095{]}, \textbf{rock} {[}0.075{]} \\
 & \templateTwo & Tom Petty : Greatest Hits is a \_\_\_\_\_ music album. & rock & country {[}0.094{]}, 2003 {[}0.084{]} \\
 & \templateThree & Tom Petty : Greatest Hits is a music album of the genre \_\_\_\_\_. & rock & \textbf{rock} {[}0.730{]}, country {[}0.104{]} \\ \bottomrule
\end{tabular}
\end{table*}



\section{Results}

In this section, we first discuss the results of the probes for genre, followed by the probes for search and recommendation. We then analyze how BERT performs in our downstream task of conversational recommendation.

\subsection{Probing BERT (RQ1)}

\subsubsection{Genres}
The results for probing BERT for each item's genre (100k books and music albums and 62k movies) are displayed in Table ~\ref{table:results_genre_probe}. We show the recall at positions 1 and 5 (number of relevant tokens in the first and first 5 predictions divided by the total number of relevant genres). To provide the reader with an intuition, we provide example prompts and predictions in Table~\ref{table:bert_cat_examples}. First we note that by just using the domain of the item, and not an item's title (\templateOne{} templates), BERT can already retrieve a reasonable amount of tokens related to the genre in the first five positions (from 25\% to 41\% depending on the domain) which is high given that the vocabulary contains ~29k tokens. We see examples of this in Table~\ref{table:bert_cat_examples}, where for instance BERT predicts \textit{fantasy} if you ask for a book genre and \textit{pop} if you ask for an album genre. This result shows that the pre-trained model indeed contains information regarding which genres are specific to each domain.

When we consider the template types where we inform BERT about the item's title (\templateTwo{} and \templateThree{}), we see that there is knowledge about specific items stored in BERT's parameters, as the results of \templateThree{} are better than \templateOne, with improvements from 0.067 to 0.179 R@1. \textbf{We can thus answer RQ1 partially: BERT has content-knowledge about items stored in its parameter, specifically regarding their genres.} From a total of ~29k tokens it can find the correct genre token up to 50\% of the times in the first 5 positions using \templateThree{}.

We also note that a prompt with more specific information leads to better results (from \templateTwo{} to \templateThree{} for instance), and this is only a lower bound for the knowledge stored, since some information might be stored in BERT that we could have retrieved with a different prompt template sentence. For example, if we do not indicate in the prompt that we are looking for the genres of the items (\templateTwo{}), we get tokens that can describe the item but are not genres. For example, for the prompt \textit{"The Wind-Up Bird Chronicle by Haruki Murakami is a \_\_\_\_\_ book."} we get the token '\textit{japanese}', (cf. Table~\ref{table:bert_cat_examples}), which is valid since the author is japanese, but it is not the correct answer for the genre probe task. Other examples when using \templateTwo retrieve for instance the publication year of the item, e.g. \textit{ "1990 book"}.



\subsubsection{Search and Recommendation}


\begin{table*}[tbp]
\centering
\footnotesize
\caption{Results for recommendation and search probes using NSP-based and SIM-based probes. Bold indicates statistical significance compared to all baselines (paired t-tests with Bonferroni correction and confidence level of 0.95). BERT stores more content-based knowledge (search) than collaborative-based knowledge (recommendation).}
\label{table:results_search_and_rec_probes}
\begin{tabular}{@{}llllllllllllll@{}}
\toprule
 &  & \multicolumn{6}{c}{Recommendation probes} & \multicolumn{6}{c}{Search probes} \\ \cmidrule(lr{1em}){3-8} \cmidrule(lr{1em}){9-14}
 &  & \multicolumn{2}{c}{Books} & \multicolumn{2}{c}{Movies} & \multicolumn{2}{c}{Music} & \multicolumn{2}{c}{Books} & \multicolumn{2}{c}{Movies} & \multicolumn{2}{c}{Music} \\ \cmidrule(lr{1em}){3-8} \cmidrule(lr{1em}){9-14}
 Technique & Model & $R_{2}@1$ & $R_{5}@1$ & $R_{2}@1$ & $R_{5}@1$ & $R_{2}@1$ & $R_{5}@1$ & $R_{2}@1$ & $R_{5}@1$ & $R_{2}@1$ & $R_{5}@1$ & $R_{2}@1$ & $R_{5}@1$ \\ \cmidrule(r{1em}){1-8} \cmidrule(l{1em}){9-14}
 - & Random & 0.500 & 0.200 & 0.500 & 0.200 & 0.500 & 0.200 & 0.500 & 0.200 & 0.500 & 0.200 & 0.500 & 0.200 \\\midrule
\multirow{1}{*}{$SIM_{CLS}$} & $BERT$ & 0.538 & 0.252 & 0.525 & 0.230 & 0.537 & 0.254 & 0.495 & 0.198 & 0.387 & 0.123 & 0.498 & 0.200 \\ 
$SIM_{CLS}$ & $RoBERTa$ & 0.574 & 0.291 & 0.509 & 0.219 & 0.550 & 0.267 & 0.578 & 0.255 & 0.516 & 0.229 & 0.527 & 0.215 \\
\multirow{1}{*}{$SIM_{MEAN}$} & $BERT$ & 0.601 & 0.331 & 0.525 & 0.232 & 0.583 & 0.295 & 0.612 & 0.338 & 0.523 & 0.235 & 0.579 & 0.314 \\ 
$SIM_{MEAN}$ & $RoBERTa$ & 0.518 & 0.230 & 0.497 & 0.205 & 0.534 & 0.243 & 0.548 & 0.225 & 0.476 & 0.208 & 0.492 & 0.192\\\hline
\multirow{1}{*}{$NSP$} & $BERT$ & \textbf{0.651} & \textbf{0.402} & \textbf{0.653} & \textbf{0.367} & \textbf{0.610} & \textbf{0.333} & \textbf{0.825} & \textbf{0.636} & \textbf{0.670 } & \textbf{0.420} & \textbf{0.755} & \textbf{0.537} \\ \bottomrule
\end{tabular}
\end{table*}


The results of the search and recommendation probes are shown in Table ~\ref{table:results_search_and_rec_probes}. We show the recall at 1 with 2 and 5 candidates $R_{2}@1$ \& $R_{5}@1$ (we resort to using different number of candidates here, due to the candidates being sentences and not tokens like the genre probing task). We see that using both \simt{} and \nsp{} techniques BERT retrieves better than the random baseline (being equal to the random baseline would mean that there is no such information stored in BERT's parameters). \textbf{This answers RQ1: BERT has content-knowledge and collaborative-knowledge about items stored in its parameter.} Using the \nsp{} technique BERT matches items with their respective reviews 82\%, 67\% and 75\% of the times for the books, movies and music domains, when choosing between two options. Also BERT selects the most appropriate item to match a user history (recommendation probe) $~$65\% of the times when, when choosing between two options.

Regarding the technique to probe BERT with, \nsp{} is the most effective, showing that this pre-training objective is indeed crucial for tasks that requires relationships between sentences. Although RoBERTa uses a similar framework to BERT, but released models with more parameters (340M → 355M), trained on more data (16GB → 160GB of text) for longer (100K → 500K steps), BERT is still more effective than RoBERTa, when we employ the \nsp{} head. We note that during the training phase of RoBERTa the NSP pre-training objective was not employed as for NLP downstream tasks no performance gains were observed~\cite{liu2019roberta}.

\begin{figure}
    \centering
    \includegraphics[width=0.40\textwidth]{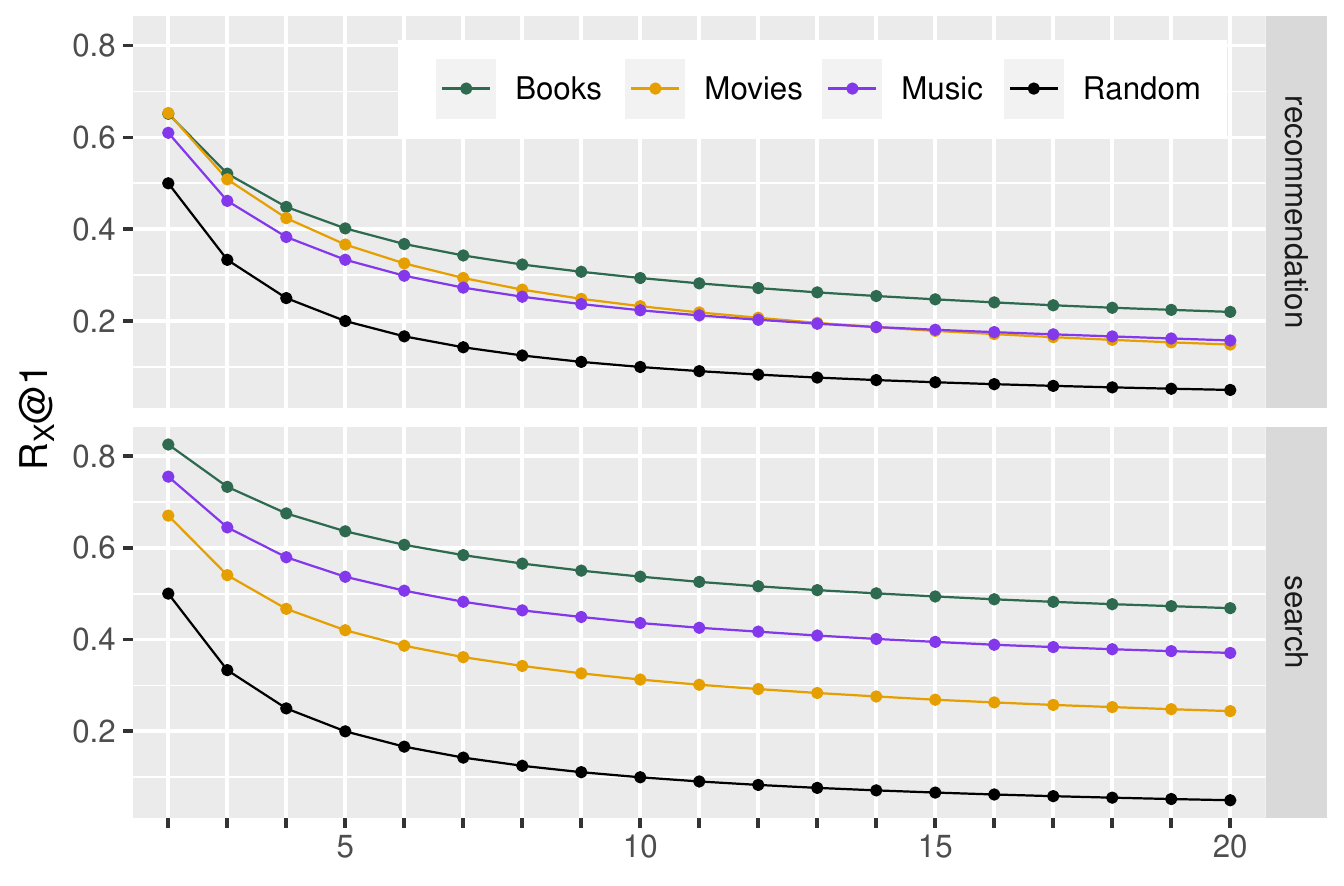}
    \setlength{\abovecaptionskip}{-1pt}
    \captionof{figure}{BERT effectiveness ($R_{x}@1$) for NSP probes when increasing the number of candidates $x$.}
    \label{fig:r_per_candidates}    
\end{figure}

We see that BERT has about 17\% more content-based knowledge (search) than collaborative-based knowledge (recommendation) considering the results from our probes. We hypothesize that this is due to textual descriptions of items with content information (useful for search) being more common than comparative sentences between different items (useful for recommendation) in the data used for BERT's pre-training. We also note in Figure~\ref{fig:r_per_candidates} that when increasing the number of candidates (x-axis), the effectiveness for the recommendation probe degrades more than for the search probes. This means that for a downstream task BERT would have to be employed as a re-ranker for only a few candidates.

When comparing different domains, the highest observed effectiveness when probing BERT for search is for books. We hypothesize this to be due to one of BERT's pre-training data being the BookCorpus~\cite{zhu2015aligning}. Since the review data used for the search probe often includes mentions of book content, the overlap between both data sources is probably high. We are unable to verify this directly because the BookCorpus dataset is not publicly available anymore. 

\subsection{Infusing knowledge into BERT for Conversational Recommendation (RQ2)}


\begin{table*}[tbp]
\centering
\footnotesize
\caption{Results for the conversational recommendation task. We provide the mean nDCG@10 and MRR, with the respective standard deviation (for 5 runs). Bold indicates statistical significance compared to all baselines (paired t-tests with Bonferroni correction and confidence level of 0.95). Fine-tuned BERT is remarkably effective for retrieving relevant answers in conversations that are about recommendations when sampling 50 negative candidates with BM25.}
\label{table:crr_results}
\begin{tabular}{@{}lllllll@{}}
\toprule
 & \multicolumn{2}{c}{/r/booksuggestions} & \multicolumn{2}{c}{/r/moviessuggestions} & \multicolumn{2}{c}{/r/musicsuggestions} \\ \cmidrule(l){2-7} 
 & nDCG@10 & MRR & nDCG@10 & MRR & nDCG@10 & MRR \\ \midrule
QL & 0.053 (.00) & 0.055 (.00) & 0.047 (.00) & 0.048 (.00) & 0.069 (.00) & 0.061 (.00) \\
RM3 & 0.050 (.00) & 0.051 (.00) & 0.043 (.00) & 0.046 (.00) & 0.052 (.00) & 0.049 (.00) \\ 
DAM & 0.549 (.01) & 0.610 (.02) & 0.716 (.02) & 0.662 (.02) & 0.238 (.04) & 0.208 (.04) \\
MSN & 0.669 (.01) & 0.707 (.01) & 0.815 (.01) & 0.788 (.02) & 0.573 (.03) & 0.535 (.06) \\ \hline
BERT & \textbf{0.903} (.01) & \textbf{0.886} (.01) & \textbf{0.938} (.00) & \textbf{0.929} (.00)  &\textbf{ 0.657} (.03)	 &  \textbf{0.620} (.03)\\ \bottomrule
\end{tabular}
\end{table*}


Table~\ref{table:crr_results} shows the results of fine-tuning BERT for the conversational recommendation task on the three domains using our Reddit forum data. Standard IR baselines, QL and QL with RM3 performed poorly on this task ($\approx$0.05 nDCG@10). We hypothesize this happens due to the recommendation nature of the underlying task in the conversation. For example, a user that describes its previously liked items does not want to receive answers with the same items being recommended in it (which are highly ranked by QL) but new item titles that have semantic similarity with the conversational context. The deep models (DMN and MSN) that learn semantic interactions between utterances and responses on the other hand perform better than traditional IR methods (from 0.5 to 0.8 nDCG@10), MSN being the best non-BERT approach. BERT is powerful at this task (up to 0.98 nDCG@10), with statistically significant improvements---35\%, 15\%, and 16\% nDCG@10 improvements for books, movies and music respectively when compared to MSN.

To investigate why BERT is so successful at this task, we resort to the ReDial dataset. Specifically, we create adversarial response candidates for the responses that included a recommendation. This is possible because unlike our Reddit-based corpus, ReDial has additional labels indicating which item from ML25M was recommended at each utterance. For every relevant response containing a recommendation, we generate adversarial candidates by changing only the relevant item with randomly selected items, see Table~\ref{table:redial_adv_examples} for some examples. This way, we can evaluate whether BERT is only picking up linguistic cues of what makes a natural response to a dialogue context or if it is using collaborative knowledge to retrieve relevant items to recommend.



The results for the adversarial dataset are displayed in Table~\ref{table:redial_results}. BERT's effectiveness drops significantly (from 0.81 nDCG@10 to 0.06 nDCG@10) when we test using the adversarial version of ReDial. Previous works have also been able to generate adversarial examples that fool BERT on different NLP tasks~\cite{jin2019bert,sun2020advbert}. 

Failing on the adversarial data shows that BERT is not able to successfully distinguish relevant items from non-relevant items, and is only using linguistic cues to find relevant answers. This motivates infusing additional knowledge into BERT, besides fine-tuning it for the conversational recommendation task. In order to do so, we do multi-task learning for the probe tasks in order to infuse additional content-based ($BERT_{search}$) and collaborative-based ($BERT_{rec}$) knowledge into BERT using only probes for items that are mentioned in the training conversations. 

\begin{table*}
\footnotesize
\caption{Examples of the ReDial dataset for conversational recommendation using either BM25 to sample negative candidates ($ReDial_{BM25}$) or the adversarial generation that replaces \textcolor{blue}{the movies} from the relevant response with random movies ($ReDial_{Adv}$) but keeps the \textcolor{purple}{context}. The adversarial candidates requires BERT to be able to chose between different movies, while for the BM25 candidates BERT can use language cues to select the correct response---likely text given the context.}
\label{table:redial_adv_examples}
\begin{tabular}{@{}P{5cm}P{3cm}P{3cm}P{3cm}@{}}
\toprule
\textbf{Context} & \textbf{Relevant response} & \textbf{Negative BM25 candidate  ($ReDial_{BM25}$)} & \textbf{Negative adversarial candidate ($ReDial_{Adv}$)} \\ \midrule
Good morning, I am looking for a good classic today. {[}SEP{]} What type of movie are you looking for today? {[}SEP{]}  I enjoyed Annie  (1982) & \textcolor{purple}{okay no problem.. If you enjoyed Annie then you will love } \textcolor{blue}{You've Got Mail (1998)} ! & I am great! What type of movie are you looking for today? & \textcolor{purple}{okay no problem.. If you enjoyed Annie then you will love} \textcolor{blue}{The Best Years of Our Lives (1946)} ! \\ \hline
HI! {[}SEP{]} Hi what type of movie would you like? {[}SEP{]} I am looking for something like Star Wars  (1977) but not Star Trek & \textcolor{purple}{Have you seen} \textcolor{blue}{Avatar  (2009)} & I love Star Trek Generations (1994)  the best! & \textcolor{purple}{Have you seen} \textcolor{blue}{Wishmaster  (1997)}, \\ \bottomrule
\end{tabular}
\end{table*}

Our results in Table~\ref{table:redial_results} show that the recommendation probe improves BERT by 9\% for the adversarial dataset $ReDial_{Adv}$, while the search probe improves effectiveness on $ReDial_{BM25}$  by 1\%. This indicates that the adversarial dataset indeed requires more collaborative-based knowledge. The approach of multi-task learning for infusing knowledge into BERT was not successful for our Reddit-based forum data. We hypothesize that this happened because, unlike ReDial, we have no additional labels indicating which items were mentioned in the reddit conversations. This forces us to train on probes for items that are likely not going to be useful. We leave the study of automatically identifying mentions to items in conversations as future work.



\begin{table}[h]
    \centering
    \caption{Fine-tuned BERT results for conversational recommendation for the dataset when using different procedures to sample negative candidates. Bold indicates statistical significance compared to other approaches (paired t-tests with Bonferroni correction and confidence level of 0.95). $BERT$ fine-tunes on ReDial, $BERT_{rec}$ multi-tasks between fine-tuning for ReDial and for the recommendation probes and $BERT_{rec}$ multi-tasks between fine-tuning for ReDial and for the search probes.}
    \label{table:redial_results}
    \begin{tabular}{@{}lllll@{}}
    \toprule
     & \multicolumn{2}{l}{$ReDial_{BM25}$} & \multicolumn{2}{l}{$ReDial_{Adv}$} \\ \cmidrule(lr{1em}){2-3}  \cmidrule(lr{1em}){4-5} 
     & nDCG@10 & MRR & nDCG@10 & MRR \\ \midrule
    $BERT$ & 0.810 (.01) & 0.778 (.01) & 0.063 (.02) & 0.069 (.02) \\ 
    $BERT_{rec}$ & 0.812 (.01) & 0.780 (.00) & \textbf{0.069 (.01)} & \textbf{0.073 (.01)} \\
    $BERT_{search}$ & \textbf{0.819 (.01) }& \textbf{0.791 (.01) }& 0.068 (.01) & 0.072 (.02) \\ \bottomrule
    \end{tabular}
    \end{table}



Answering our second research question \textbf{(RQ2), we demonstrate that infusing knowledge from the probing tasks into BERT, via multi-task learning during the fine-tuning procedure is an effective technique}, with improvements of up to 9\% of nDCG@10 for conversational recommendation.



\section{Conclusions}
Given the potential that heavily pre-trained LMs offer for conversational recommender systems, in this paper we examine how much knowledge is stored in BERT's parameters regarding books, movies and music. We resort to different probes in order to answer this question. We find that we can use BERT to extract the genre for 30-50\% of the items on the top 5 predictions, depending on the domain; and that BERT has about 17\% more content-based knowledge (search) than collaborative-based knowledge (recommendation).

Based on the findings of our probing task we investigate a retrieval-based approach based on BERT for conversational recommendation, and how to infuse knowledge into its parameters. We show that BERT is powerful for distinguishing relevant from non-relevant responses (0.9 nDCG@10 compared to the second best baseline with 0.7 nDCG@10). By using adversarial data, we demonstrate that BERT is less effective when it has to distinguish candidate responses that are reasonable responses but include randomly selected item recommendations. This motivates infusing collaborative-based and content-based knowledge in the probing tasks into BERT, which we do via multi-task learning during the fine-tuning step and show effectiveness improvements of up to 9\%.

Overall, we provide insights on what BERT can do with the knowledge it has stored in its parameters that can be helpful to build CRS, where it fails and how we can infuse knowledge into BERT. As future work, user studies are important to investigate to what extent a BERT based model works in practice for conversational recommendation. Extending our work for other domains can also provide more insights on the utility of BERT for CRS. 


\begin{acks}
This research has been supported by NWO projects SearchX (639.022.722) and NWO Aspasia (015.013.027).
\end{acks}

\bibliographystyle{ACM-Reference-Format}





\begin{thebibliography}{63}


\ifx \showCODEN    \undefined \def \showCODEN     #1{\unskip}     \fi
\ifx \showDOI      \undefined \def \showDOI       #1{#1}\fi
\ifx \showISBNx    \undefined \def \showISBNx     #1{\unskip}     \fi
\ifx \showISBNxiii \undefined \def \showISBNxiii  #1{\unskip}     \fi
\ifx \showISSN     \undefined \def \showISSN      #1{\unskip}     \fi
\ifx \showLCCN     \undefined \def \showLCCN      #1{\unskip}     \fi
\ifx \shownote     \undefined \def \shownote      #1{#1}          \fi
\ifx \showarticletitle \undefined \def \showarticletitle #1{#1}   \fi
\ifx \showURL      \undefined \def \showURL       {\relax}        \fi
\providecommand\bibfield[2]{#2}
\providecommand\bibinfo[2]{#2}
\providecommand\natexlab[1]{#1}
\providecommand\showeprint[2][]{arXiv:#2}

\bibitem[\protect\citeauthoryear{Ai, Zhang, Bi, Chen, and Croft}{Ai
  et~al\mbox{.}}{2017}]%
        {ai2017learning}
\bibfield{author}{\bibinfo{person}{Qingyao Ai}, \bibinfo{person}{Yongfeng
  Zhang}, \bibinfo{person}{Keping Bi}, \bibinfo{person}{Xu Chen}, {and}
  \bibinfo{person}{W~Bruce Croft}.} \bibinfo{year}{2017}\natexlab{}.
\newblock \showarticletitle{Learning a hierarchical embedding model for
  personalized product search}. In \bibinfo{booktitle}{\emph{SIGIR}}.
  \bibinfo{pages}{645--654}.
\newblock


\bibitem[\protect\citeauthoryear{Azzopardi, Dubiel, Halvey, and
  Dalton}{Azzopardi et~al\mbox{.}}{2018}]%
        {azzopardi2018conceptualizing}
\bibfield{author}{\bibinfo{person}{Leif Azzopardi}, \bibinfo{person}{Mateusz
  Dubiel}, \bibinfo{person}{Martin Halvey}, {and} \bibinfo{person}{Jeffery
  Dalton}.} \bibinfo{year}{2018}\natexlab{}.
\newblock \showarticletitle{Conceptualizing agent-human interactions during the
  conversational search process}. In \bibinfo{booktitle}{\emph{CAIR}}.
\newblock


\bibitem[\protect\citeauthoryear{Bellogin, Castells, and Cantador}{Bellogin
  et~al\mbox{.}}{2011}]%
        {10.1145/2043932.2043996}
\bibfield{author}{\bibinfo{person}{Alejandro Bellogin}, \bibinfo{person}{Pablo
  Castells}, {and} \bibinfo{person}{Ivan Cantador}.}
  \bibinfo{year}{2011}\natexlab{}.
\newblock \showarticletitle{Precision-Oriented Evaluation of Recommender
  Systems: An Algorithmic Comparison}. In \bibinfo{booktitle}{\emph{RecSys}}.
  \bibinfo{pages}{333–336}.
\newblock
\showISBNx{9781450306836}


\bibitem[\protect\citeauthoryear{C{\^a}mara and Hauff}{C{\^a}mara and
  Hauff}{2020}]%
        {camara2020diagnosing}
\bibfield{author}{\bibinfo{person}{Arthur C{\^a}mara} {and}
  \bibinfo{person}{Claudia Hauff}.} \bibinfo{year}{2020}\natexlab{}.
\newblock \showarticletitle{Diagnosing BERT with Retrieval Heuristics}. In
  \bibinfo{booktitle}{\emph{ECIR}}. Springer, \bibinfo{pages}{605--618}.
\newblock


\bibitem[\protect\citeauthoryear{Clark, Khandelwal, Levy, and Manning}{Clark
  et~al\mbox{.}}{2019}]%
        {clark2019does}
\bibfield{author}{\bibinfo{person}{Kevin Clark}, \bibinfo{person}{Urvashi
  Khandelwal}, \bibinfo{person}{Omer Levy}, {and}
  \bibinfo{person}{Christopher~D Manning}.} \bibinfo{year}{2019}\natexlab{}.
\newblock \showarticletitle{What Does BERT Look at? An Analysis of BERT’s
  Attention}. In \bibinfo{booktitle}{\emph{ACL Workshop BlackboxNLP}}.
  \bibinfo{pages}{276--286}.
\newblock


\bibitem[\protect\citeauthoryear{Devlin, Chang, Lee, and Toutanova}{Devlin
  et~al\mbox{.}}{2019}]%
        {devlin2018bert}
\bibfield{author}{\bibinfo{person}{Jacob Devlin}, \bibinfo{person}{Ming-Wei
  Chang}, \bibinfo{person}{Kenton Lee}, {and} \bibinfo{person}{Kristina
  Toutanova}.} \bibinfo{year}{2019}\natexlab{}.
\newblock \showarticletitle{BERT: Pre-training of Deep Bidirectional
  Transformers for Language Understanding}. In
  \bibinfo{booktitle}{\emph{NAACL-HLT}}.
\newblock


\bibitem[\protect\citeauthoryear{Guo, Cheng, Nie, Wang, Ma, and
  Kankanhalli}{Guo et~al\mbox{.}}{2019}]%
        {guo2019attentive}
\bibfield{author}{\bibinfo{person}{Yangyang Guo}, \bibinfo{person}{Zhiyong
  Cheng}, \bibinfo{person}{Liqiang Nie}, \bibinfo{person}{Yinglong Wang},
  \bibinfo{person}{Jun Ma}, {and} \bibinfo{person}{Mohan Kankanhalli}.}
  \bibinfo{year}{2019}\natexlab{}.
\newblock \showarticletitle{Attentive long short-term preference modeling for
  personalized product search}.
\newblock \bibinfo{journal}{\emph{TOIS}} \bibinfo{volume}{37},
  \bibinfo{number}{2} (\bibinfo{year}{2019}), \bibinfo{pages}{1--27}.
\newblock


\bibitem[\protect\citeauthoryear{Gururangan, Marasovi{\'c}, Swayamdipta, Lo,
  Beltagy, Downey, and Smith}{Gururangan et~al\mbox{.}}{2020}]%
        {gururangan2020don}
\bibfield{author}{\bibinfo{person}{Suchin Gururangan}, \bibinfo{person}{Ana
  Marasovi{\'c}}, \bibinfo{person}{Swabha Swayamdipta}, \bibinfo{person}{Kyle
  Lo}, \bibinfo{person}{Iz Beltagy}, \bibinfo{person}{Doug Downey}, {and}
  \bibinfo{person}{Noah~A. Smith}.} \bibinfo{year}{2020}\natexlab{}.
\newblock \showarticletitle{Don{'}t Stop Pretraining: Adapt Language Models to
  Domains and Tasks}. In \bibinfo{booktitle}{\emph{ACL}}.
  \bibinfo{pages}{8342--8360}.
\newblock


\bibitem[\protect\citeauthoryear{Harper and Konstan}{Harper and
  Konstan}{2015}]%
        {harper2015movielens}
\bibfield{author}{\bibinfo{person}{F~Maxwell Harper} {and}
  \bibinfo{person}{Joseph~A Konstan}.} \bibinfo{year}{2015}\natexlab{}.
\newblock \showarticletitle{The movielens datasets: History and context}.
\newblock \bibinfo{journal}{\emph{TiiS}} \bibinfo{volume}{5},
  \bibinfo{number}{4} (\bibinfo{year}{2015}), \bibinfo{pages}{1--19}.
\newblock


\bibitem[\protect\citeauthoryear{Jannach, Manzoor, Cai, and Chen}{Jannach
  et~al\mbox{.}}{2020}]%
        {jannach2020survey}
\bibfield{author}{\bibinfo{person}{Dietmar Jannach}, \bibinfo{person}{Ahtsham
  Manzoor}, \bibinfo{person}{Wanling Cai}, {and} \bibinfo{person}{Li Chen}.}
  \bibinfo{year}{2020}\natexlab{}.
\newblock \showarticletitle{A Survey on Conversational Recommender Systems}.
\newblock \bibinfo{journal}{\emph{arXiv:2004.00646}} (\bibinfo{year}{2020}).
\newblock


\bibitem[\protect\citeauthoryear{Jawahar, Sagot, and Seddah}{Jawahar
  et~al\mbox{.}}{2019}]%
        {jawahar2019does}
\bibfield{author}{\bibinfo{person}{Ganesh Jawahar},
  \bibinfo{person}{Beno{\^\i}t Sagot}, {and} \bibinfo{person}{Djam{\'e}
  Seddah}.} \bibinfo{year}{2019}\natexlab{}.
\newblock \showarticletitle{What does BERT learn about the structure of
  language?}. In \bibinfo{booktitle}{\emph{ACL}}.
\newblock


\bibitem[\protect\citeauthoryear{Jiang, Xu, Araki, and Neubig}{Jiang
  et~al\mbox{.}}{2020}]%
        {jiang2019can}
\bibfield{author}{\bibinfo{person}{Zhengbao Jiang}, \bibinfo{person}{Frank~F
  Xu}, \bibinfo{person}{Jun Araki}, {and} \bibinfo{person}{Graham Neubig}.}
  \bibinfo{year}{2020}\natexlab{}.
\newblock \showarticletitle{How Can We Know What Language Models Know?}
\newblock \bibinfo{journal}{\emph{TACL}}  \bibinfo{volume}{8}
  (\bibinfo{year}{2020}), \bibinfo{pages}{423--438}.
\newblock


\bibitem[\protect\citeauthoryear{Jin, Jin, Zhou, and Szolovits}{Jin
  et~al\mbox{.}}{2020}]%
        {jin2019bert}
\bibfield{author}{\bibinfo{person}{Di Jin}, \bibinfo{person}{Zhijing Jin},
  \bibinfo{person}{Joey~Tianyi Zhou}, {and} \bibinfo{person}{Peter Szolovits}.}
  \bibinfo{year}{2020}\natexlab{}.
\newblock \showarticletitle{Is BERT Really Robust? A Strong Baseline for
  Natural Language Attack on Text Classification and Entailment}. In
  \bibinfo{booktitle}{\emph{AAAI}}.
\newblock


\bibitem[\protect\citeauthoryear{Kendall, Gal, and Cipolla}{Kendall
  et~al\mbox{.}}{2018}]%
        {Kendall_2018_CVPR}
\bibfield{author}{\bibinfo{person}{Alex Kendall}, \bibinfo{person}{Yarin Gal},
  {and} \bibinfo{person}{Roberto Cipolla}.} \bibinfo{year}{2018}\natexlab{}.
\newblock \showarticletitle{Multi-Task Learning Using Uncertainty to Weigh
  Losses for Scene Geometry and Semantics}. In
  \bibinfo{booktitle}{\emph{CVPR}}.
\newblock


\bibitem[\protect\citeauthoryear{Kingma and Ba}{Kingma and Ba}{2014}]%
        {kingma2014adam}
\bibfield{author}{\bibinfo{person}{Diederik Kingma} {and}
  \bibinfo{person}{Jimmy Ba}.} \bibinfo{year}{2014}\natexlab{}.
\newblock \showarticletitle{Adam: A Method for Stochastic Optimization}.
\newblock \bibinfo{journal}{\emph{ICLR}} (\bibinfo{date}{12}
  \bibinfo{year}{2014}).
\newblock


\bibitem[\protect\citeauthoryear{Kirkpatrick, Pascanu, Rabinowitz, Veness,
  Desjardins, Rusu, Milan, Quan, Ramalho, Grabska-Barwinska,
  et~al\mbox{.}}{Kirkpatrick et~al\mbox{.}}{2017}]%
        {kirkpatrick2017overcoming}
\bibfield{author}{\bibinfo{person}{James Kirkpatrick}, \bibinfo{person}{Razvan
  Pascanu}, \bibinfo{person}{Neil Rabinowitz}, \bibinfo{person}{Joel Veness},
  \bibinfo{person}{Guillaume Desjardins}, \bibinfo{person}{Andrei~A Rusu},
  \bibinfo{person}{Kieran Milan}, \bibinfo{person}{John Quan},
  \bibinfo{person}{Tiago Ramalho}, \bibinfo{person}{Agnieszka
  Grabska-Barwinska}, {et~al\mbox{.}}} \bibinfo{year}{2017}\natexlab{}.
\newblock \showarticletitle{Overcoming catastrophic forgetting in neural
  networks}.
\newblock \bibinfo{journal}{\emph{Proceedings of the national academy of
  sciences}} \bibinfo{volume}{114}, \bibinfo{number}{13}
  (\bibinfo{year}{2017}), \bibinfo{pages}{3521--3526}.
\newblock


\bibitem[\protect\citeauthoryear{Lavrenko and Croft}{Lavrenko and
  Croft}{2001}]%
        {10.1145/383952.383972}
\bibfield{author}{\bibinfo{person}{Victor Lavrenko} {and}
  \bibinfo{person}{W.~Bruce Croft}.} \bibinfo{year}{2001}\natexlab{}.
\newblock \showarticletitle{Relevance Based Language Models}. In
  \bibinfo{booktitle}{\emph{SIGIR}}. \bibinfo{pages}{120–127}.
\newblock
\showISBNx{1581133316}


\bibitem[\protect\citeauthoryear{Li, Kahou, Schulz, Michalski, Charlin, and
  Pal}{Li et~al\mbox{.}}{2018}]%
        {li2018towards}
\bibfield{author}{\bibinfo{person}{Raymond Li},
  \bibinfo{person}{Samira~Ebrahimi Kahou}, \bibinfo{person}{Hannes Schulz},
  \bibinfo{person}{Vincent Michalski}, \bibinfo{person}{Laurent Charlin}, {and}
  \bibinfo{person}{Chris Pal}.} \bibinfo{year}{2018}\natexlab{}.
\newblock \showarticletitle{Towards deep conversational recommendations}. In
  \bibinfo{booktitle}{\emph{NeurIPS}}. \bibinfo{pages}{9725--9735}.
\newblock


\bibitem[\protect\citeauthoryear{Liu, Ott, Goyal, Du, Joshi, Chen, Levy, Lewis,
  Zettlemoyer, and Stoyanov}{Liu et~al\mbox{.}}{2019}]%
        {liu2019roberta}
\bibfield{author}{\bibinfo{person}{Yinhan Liu}, \bibinfo{person}{Myle Ott},
  \bibinfo{person}{Naman Goyal}, \bibinfo{person}{Jingfei Du},
  \bibinfo{person}{Mandar Joshi}, \bibinfo{person}{Danqi Chen},
  \bibinfo{person}{Omer Levy}, \bibinfo{person}{Mike Lewis},
  \bibinfo{person}{Luke Zettlemoyer}, {and} \bibinfo{person}{Veselin
  Stoyanov}.} \bibinfo{year}{2019}\natexlab{}.
\newblock \showarticletitle{Roberta: A robustly optimized bert pretraining
  approach}.
\newblock \bibinfo{journal}{\emph{arXiv:1907.11692}} (\bibinfo{year}{2019}).
\newblock


\bibitem[\protect\citeauthoryear{Michel, Levy, and Neubig}{Michel
  et~al\mbox{.}}{2019}]%
        {michel2019sixteen}
\bibfield{author}{\bibinfo{person}{Paul Michel}, \bibinfo{person}{Omer Levy},
  {and} \bibinfo{person}{Graham Neubig}.} \bibinfo{year}{2019}\natexlab{}.
\newblock \showarticletitle{Are Sixteen Heads Really Better than One?}. In
  \bibinfo{booktitle}{\emph{NeurIPS}}. \bibinfo{pages}{14014--14024}.
\newblock


\bibitem[\protect\citeauthoryear{Ni, Li, and McAuley}{Ni et~al\mbox{.}}{2019}]%
        {ni2019justifying}
\bibfield{author}{\bibinfo{person}{Jianmo Ni}, \bibinfo{person}{Jiacheng Li},
  {and} \bibinfo{person}{Julian McAuley}.} \bibinfo{year}{2019}\natexlab{}.
\newblock \showarticletitle{Justifying Recommendations using Distantly-Labeled
  Reviews and Fine-Grained Aspects}. In
  \bibinfo{booktitle}{\emph{EMNLP-IJCNLP}}. \bibinfo{pages}{188--197}.
\newblock


\bibitem[\protect\citeauthoryear{Oddy}{Oddy}{1977}]%
        {oddy1977information}
\bibfield{author}{\bibinfo{person}{Robert~N Oddy}.}
  \bibinfo{year}{1977}\natexlab{}.
\newblock \showarticletitle{Information retrieval through man-machine
  dialogue}.
\newblock \bibinfo{journal}{\emph{Journal of documentation}}
  \bibinfo{volume}{33}, \bibinfo{number}{1} (\bibinfo{year}{1977}),
  \bibinfo{pages}{1--14}.
\newblock


\bibitem[\protect\citeauthoryear{Penha, Balan, and Hauff}{Penha
  et~al\mbox{.}}{2019}]%
        {penha2019introducing}
\bibfield{author}{\bibinfo{person}{Gustavo Penha}, \bibinfo{person}{Alexandru
  Balan}, {and} \bibinfo{person}{Claudia Hauff}.}
  \bibinfo{year}{2019}\natexlab{}.
\newblock \bibinfo{title}{Introducing MANtIS: a novel Multi-Domain Information
  Seeking Dialogues Dataset}.
\newblock
\newblock
\showeprint[arxiv]{1912.04639}


\bibitem[\protect\citeauthoryear{Penha and Hauff}{Penha and Hauff}{2020}]%
        {penha2020curriculum}
\bibfield{author}{\bibinfo{person}{Gustavo Penha} {and}
  \bibinfo{person}{Claudia Hauff}.} \bibinfo{year}{2020}\natexlab{}.
\newblock \showarticletitle{Curriculum Learning Strategies for IR}. In
  \bibinfo{booktitle}{\emph{ECIR}}. Springer, \bibinfo{pages}{699--713}.
\newblock


\bibitem[\protect\citeauthoryear{Pentina, Sharmanska, and Lampert}{Pentina
  et~al\mbox{.}}{2015}]%
        {pentina2015curriculum}
\bibfield{author}{\bibinfo{person}{Anastasia Pentina},
  \bibinfo{person}{Viktoriia Sharmanska}, {and} \bibinfo{person}{Christoph~H
  Lampert}.} \bibinfo{year}{2015}\natexlab{}.
\newblock \showarticletitle{Curriculum learning of multiple tasks}. In
  \bibinfo{booktitle}{\emph{CVPR}}. \bibinfo{pages}{5492--5500}.
\newblock


\bibitem[\protect\citeauthoryear{Petroni, Lewis, Piktus, Rockt{\"a}schel, Wu,
  Miller, and Riedel}{Petroni et~al\mbox{.}}{2020}]%
        {petroni2020context}
\bibfield{author}{\bibinfo{person}{Fabio Petroni}, \bibinfo{person}{Patrick
  Lewis}, \bibinfo{person}{Aleksandra Piktus}, \bibinfo{person}{Tim
  Rockt{\"a}schel}, \bibinfo{person}{Yuxiang Wu}, \bibinfo{person}{Alexander~H.
  Miller}, {and} \bibinfo{person}{Sebastian Riedel}.}
  \bibinfo{year}{2020}\natexlab{}.
\newblock \showarticletitle{How Context Affects Language Models' Factual
  Predictions}. In \bibinfo{booktitle}{\emph{AKBC}}.
\newblock


\bibitem[\protect\citeauthoryear{Petroni, Rockt{\"a}schel, Riedel, Lewis,
  Bakhtin, Wu, and Miller}{Petroni et~al\mbox{.}}{2019}]%
        {petroni2019language}
\bibfield{author}{\bibinfo{person}{Fabio Petroni}, \bibinfo{person}{Tim
  Rockt{\"a}schel}, \bibinfo{person}{Sebastian Riedel},
  \bibinfo{person}{Patrick Lewis}, \bibinfo{person}{Anton Bakhtin},
  \bibinfo{person}{Yuxiang Wu}, {and} \bibinfo{person}{Alexander Miller}.}
  \bibinfo{year}{2019}\natexlab{}.
\newblock \showarticletitle{Language Models as Knowledge Bases?}. In
  \bibinfo{booktitle}{\emph{EMNLP-IJCNLP}}. \bibinfo{pages}{2463--2473}.
\newblock


\bibitem[\protect\citeauthoryear{Phang, F{\'e}vry, and Bowman}{Phang
  et~al\mbox{.}}{2018}]%
        {phang2018sentence}
\bibfield{author}{\bibinfo{person}{Jason Phang}, \bibinfo{person}{Thibault
  F{\'e}vry}, {and} \bibinfo{person}{Samuel~R Bowman}.}
  \bibinfo{year}{2018}\natexlab{}.
\newblock \showarticletitle{Sentence encoders on stilts: Supplementary training
  on intermediate labeled-data tasks}.
\newblock \bibinfo{journal}{\emph{arXiv:1811.01088}} (\bibinfo{year}{2018}).
\newblock


\bibitem[\protect\citeauthoryear{Ponte and Croft}{Ponte and Croft}{1998}]%
        {10.1145/290941.291008}
\bibfield{author}{\bibinfo{person}{Jay~M. Ponte} {and}
  \bibinfo{person}{W.~Bruce Croft}.} \bibinfo{year}{1998}\natexlab{}.
\newblock \showarticletitle{A Language Modeling Approach to Information
  Retrieval}. In \bibinfo{booktitle}{\emph{SIGIR}}
  \emph{(\bibinfo{series}{SIGIR ’98})}. \bibinfo{pages}{275–281}.
\newblock
\showISBNx{1581130155}


\bibitem[\protect\citeauthoryear{Pruksachatkun, Phang, Liu, Htut, Zhang, Pang,
  Vania, Kann, and Bowman}{Pruksachatkun et~al\mbox{.}}{2020}]%
        {pruksachatkun2020intermediate}
\bibfield{author}{\bibinfo{person}{Yada Pruksachatkun}, \bibinfo{person}{Jason
  Phang}, \bibinfo{person}{Haokun Liu}, \bibinfo{person}{Phu~Mon Htut},
  \bibinfo{person}{Xiaoyi Zhang}, \bibinfo{person}{Richard~Yuanzhe Pang},
  \bibinfo{person}{Clara Vania}, \bibinfo{person}{Katharina Kann}, {and}
  \bibinfo{person}{Samuel~R. Bowman}.} \bibinfo{year}{2020}\natexlab{}.
\newblock \showarticletitle{Intermediate-Task Transfer Learning with Pretrained
  Language Models: When and Why Does It Work?}. In
  \bibinfo{booktitle}{\emph{ACL}}. \bibinfo{pages}{5231--5247}.
\newblock


\bibitem[\protect\citeauthoryear{Qiu, Sun, Xu, Shao, Dai, and Huang}{Qiu
  et~al\mbox{.}}{2020}]%
        {qiu2020pre}
\bibfield{author}{\bibinfo{person}{Xipeng Qiu}, \bibinfo{person}{Tianxiang
  Sun}, \bibinfo{person}{Yige Xu}, \bibinfo{person}{Yunfan Shao},
  \bibinfo{person}{Ning Dai}, {and} \bibinfo{person}{Xuanjing Huang}.}
  \bibinfo{year}{2020}\natexlab{}.
\newblock \showarticletitle{Pre-trained Models for Natural Language Processing:
  A Survey}.
\newblock \bibinfo{journal}{\emph{arXiv:2003.08271}} (\bibinfo{year}{2020}).
\newblock


\bibitem[\protect\citeauthoryear{Qu, Yang, Croft, Trippas, Zhang, and Qiu}{Qu
  et~al\mbox{.}}{2018}]%
        {qu2018analyzing}
\bibfield{author}{\bibinfo{person}{Chen Qu}, \bibinfo{person}{Liu Yang},
  \bibinfo{person}{W~Bruce Croft}, \bibinfo{person}{Johanne~R Trippas},
  \bibinfo{person}{Yongfeng Zhang}, {and} \bibinfo{person}{Minghui Qiu}.}
  \bibinfo{year}{2018}\natexlab{}.
\newblock \showarticletitle{Analyzing and Characterizing User Intent in
  Information-seeking Conversations}.
\newblock \bibinfo{journal}{\emph{SIGIR}} (\bibinfo{year}{2018}).
\newblock


\bibitem[\protect\citeauthoryear{Qu, Yang, Qiu, Croft, Zhang, and Iyyer}{Qu
  et~al\mbox{.}}{2019}]%
        {Qu:2019:BHA:3331184.3331341}
\bibfield{author}{\bibinfo{person}{Chen Qu}, \bibinfo{person}{Liu Yang},
  \bibinfo{person}{Minghui Qiu}, \bibinfo{person}{W.~Bruce Croft},
  \bibinfo{person}{Yongfeng Zhang}, {and} \bibinfo{person}{Mohit Iyyer}.}
  \bibinfo{year}{2019}\natexlab{}.
\newblock \showarticletitle{{BERT with History Answer Embedding for
  Conversational Question Answering}} \emph{(\bibinfo{series}{SIGIR})}.
  \bibinfo{pages}{1133--1136}.
\newblock
\showISBNx{978-1-4503-6172-9}


\bibitem[\protect\citeauthoryear{Radlinski, Balog, Byrne, and
  Krishnamoorthi}{Radlinski et~al\mbox{.}}{2019}]%
        {radlinksi2019coached}
\bibfield{author}{\bibinfo{person}{Filip Radlinski}, \bibinfo{person}{Krisztian
  Balog}, \bibinfo{person}{Bill Byrne}, {and} \bibinfo{person}{Karthik
  Krishnamoorthi}.} \bibinfo{year}{2019}\natexlab{}.
\newblock \showarticletitle{Coached Conversational Preference Elicitation: A
  Case Study in Understanding Movie Preferences}. In
  \bibinfo{booktitle}{\emph{SIGDIAL}}.
\newblock


\bibitem[\protect\citeauthoryear{Radlinski and Craswell}{Radlinski and
  Craswell}{2017}]%
        {radlinski2017theoretical}
\bibfield{author}{\bibinfo{person}{Filip Radlinski} {and} \bibinfo{person}{Nick
  Craswell}.} \bibinfo{year}{2017}\natexlab{}.
\newblock \showarticletitle{A theoretical framework for conversational search}.
  In \bibinfo{booktitle}{\emph{CHIIR}}. \bibinfo{pages}{117--126}.
\newblock


\bibitem[\protect\citeauthoryear{Raffel, Shazeer, Roberts, Lee, Narang, Matena,
  Zhou, Li, and Liu}{Raffel et~al\mbox{.}}{2020}]%
        {raffel2019exploring}
\bibfield{author}{\bibinfo{person}{Colin Raffel}, \bibinfo{person}{Noam
  Shazeer}, \bibinfo{person}{Adam Roberts}, \bibinfo{person}{Katherine Lee},
  \bibinfo{person}{Sharan Narang}, \bibinfo{person}{Michael Matena},
  \bibinfo{person}{Yanqi Zhou}, \bibinfo{person}{Wei Li}, {and}
  \bibinfo{person}{Peter~J. Liu}.} \bibinfo{year}{2020}\natexlab{}.
\newblock \showarticletitle{Exploring the Limits of Transfer Learning with a
  Unified Text-to-Text Transformer}.
\newblock \bibinfo{journal}{\emph{JMLR}} \bibinfo{volume}{21},
  \bibinfo{number}{140} (\bibinfo{year}{2020}), \bibinfo{pages}{1--67}.
\newblock


\bibitem[\protect\citeauthoryear{Roberts, Raffel, and Shazeer}{Roberts
  et~al\mbox{.}}{2020}]%
        {roberts2020much}
\bibfield{author}{\bibinfo{person}{Adam Roberts}, \bibinfo{person}{Colin
  Raffel}, {and} \bibinfo{person}{Noam Shazeer}.}
  \bibinfo{year}{2020}\natexlab{}.
\newblock \showarticletitle{How Much Knowledge Can You Pack Into the Parameters
  of a Language Model?}
\newblock \bibinfo{journal}{\emph{arXiv:2002.08910}} (\bibinfo{year}{2020}).
\newblock


\bibitem[\protect\citeauthoryear{Robertson and Zaragoza}{Robertson and
  Zaragoza}{2009}]%
        {10.1561/1500000019}
\bibfield{author}{\bibinfo{person}{Stephen Robertson} {and}
  \bibinfo{person}{Hugo Zaragoza}.} \bibinfo{year}{2009}\natexlab{}.
\newblock \showarticletitle{The Probabilistic Relevance Framework: BM25 and
  Beyond}.
\newblock \bibinfo{journal}{\emph{Found. Trends Inf. Retr.}}
  \bibinfo{volume}{3}, \bibinfo{number}{4} (\bibinfo{date}{April}
  \bibinfo{year}{2009}), \bibinfo{pages}{333–389}.
\newblock
\showISSN{1554-0669}


\bibitem[\protect\citeauthoryear{Rosset, Xiong, Song, Campos, Craswell, Tiwary,
  and Bennett}{Rosset et~al\mbox{.}}{2020}]%
        {rosset2020leading}
\bibfield{author}{\bibinfo{person}{Corbin Rosset}, \bibinfo{person}{Chenyan
  Xiong}, \bibinfo{person}{Xia Song}, \bibinfo{person}{Daniel Campos},
  \bibinfo{person}{Nick Craswell}, \bibinfo{person}{Saurabh Tiwary}, {and}
  \bibinfo{person}{Paul Bennett}.} \bibinfo{year}{2020}\natexlab{}.
\newblock \showarticletitle{Leading Conversational Search by Suggesting Useful
  Questions}. In \bibinfo{booktitle}{\emph{WWW}}. \bibinfo{pages}{1160--1170}.
\newblock


\bibitem[\protect\citeauthoryear{Sakata, Shibata, Tanaka, and Kurohashi}{Sakata
  et~al\mbox{.}}{2019}]%
        {sakata2019faq}
\bibfield{author}{\bibinfo{person}{Wataru Sakata}, \bibinfo{person}{Tomohide
  Shibata}, \bibinfo{person}{Ribeka Tanaka}, {and} \bibinfo{person}{Sadao
  Kurohashi}.} \bibinfo{year}{2019}\natexlab{}.
\newblock \showarticletitle{FAQ retrieval using query-question similarity and
  BERT-based query-answer relevance}. In \bibinfo{booktitle}{\emph{SIGIR}}.
  \bibinfo{pages}{1113--1116}.
\newblock


\bibitem[\protect\citeauthoryear{Sun, Hashimoto, Yin, Asai, Li, Yu, and
  Xiong}{Sun et~al\mbox{.}}{2020}]%
        {sun2020advbert}
\bibfield{author}{\bibinfo{person}{Lichao Sun}, \bibinfo{person}{Kazuma
  Hashimoto}, \bibinfo{person}{Wenpeng Yin}, \bibinfo{person}{Akari Asai},
  \bibinfo{person}{Jia Li}, \bibinfo{person}{Philip Yu}, {and}
  \bibinfo{person}{Caiming Xiong}.} \bibinfo{year}{2020}\natexlab{}.
\newblock \bibinfo{title}{Adv-BERT: BERT is not robust on misspellings!
  Generating nature adversarial samples on BERT}.
\newblock
\newblock
\showeprint[arxiv]{2003.04985}


\bibitem[\protect\citeauthoryear{Talmor, Elazar, Goldberg, and Berant}{Talmor
  et~al\mbox{.}}{2019}]%
        {talmor2019olmpics}
\bibfield{author}{\bibinfo{person}{Alon Talmor}, \bibinfo{person}{Yanai
  Elazar}, \bibinfo{person}{Yoav Goldberg}, {and} \bibinfo{person}{Jonathan
  Berant}.} \bibinfo{year}{2019}\natexlab{}.
\newblock \showarticletitle{oLMpics--On what Language Model Pre-training
  Captures}.
\newblock \bibinfo{journal}{\emph{arXiv:1912.13283}} (\bibinfo{year}{2019}).
\newblock


\bibitem[\protect\citeauthoryear{Tao, Wu, Xu, Hu, Zhao, and Yan}{Tao
  et~al\mbox{.}}{2019}]%
        {tao2019one}
\bibfield{author}{\bibinfo{person}{Chongyang Tao}, \bibinfo{person}{Wei Wu},
  \bibinfo{person}{Can Xu}, \bibinfo{person}{Wenpeng Hu},
  \bibinfo{person}{Dongyan Zhao}, {and} \bibinfo{person}{Rui Yan}.}
  \bibinfo{year}{2019}\natexlab{}.
\newblock \showarticletitle{{One Time of Interaction May Not Be Enough: Go Deep
  with an Interaction-over-Interaction Network for Response Selection in
  Dialogues}}. In \bibinfo{booktitle}{\emph{ACL}}. \bibinfo{pages}{1--11}.
\newblock


\bibitem[\protect\citeauthoryear{Tenney, Das, and Pavlick}{Tenney
  et~al\mbox{.}}{2019}]%
        {tenney2019bert}
\bibfield{author}{\bibinfo{person}{Ian Tenney}, \bibinfo{person}{Dipanjan Das},
  {and} \bibinfo{person}{Ellie Pavlick}.} \bibinfo{year}{2019}\natexlab{}.
\newblock \showarticletitle{BERT Rediscovers the Classical NLP Pipeline}. In
  \bibinfo{booktitle}{\emph{ACL}}. \bibinfo{pages}{4593--4601}.
\newblock


\bibitem[\protect\citeauthoryear{Thompson, Gwinnup, Khayrallah, Duh, and
  Koehn}{Thompson et~al\mbox{.}}{2019}]%
        {thompson2019overcoming}
\bibfield{author}{\bibinfo{person}{Brian Thompson}, \bibinfo{person}{Jeremy
  Gwinnup}, \bibinfo{person}{Huda Khayrallah}, \bibinfo{person}{Kevin Duh},
  {and} \bibinfo{person}{Philipp Koehn}.} \bibinfo{year}{2019}\natexlab{}.
\newblock \showarticletitle{Overcoming catastrophic forgetting during domain
  adaptation of neural machine translation}. In
  \bibinfo{booktitle}{\emph{NAACL}}. \bibinfo{pages}{2062--2068}.
\newblock


\bibitem[\protect\citeauthoryear{Trippas, Spina, Cavedon, Joho, and
  Sanderson}{Trippas et~al\mbox{.}}{2018}]%
        {trippas2018informing}
\bibfield{author}{\bibinfo{person}{Johanne~R Trippas}, \bibinfo{person}{Damiano
  Spina}, \bibinfo{person}{Lawrence Cavedon}, \bibinfo{person}{Hideo Joho},
  {and} \bibinfo{person}{Mark Sanderson}.} \bibinfo{year}{2018}\natexlab{}.
\newblock \showarticletitle{Informing the Design of Spoken Conversational
  Search: Perspective Paper}. In \bibinfo{booktitle}{\emph{CHIIR}}.
  \bibinfo{pages}{32--41}.
\newblock


\bibitem[\protect\citeauthoryear{Vakulenko, Revoredo, Di~Ciccio, and
  de~Rijke}{Vakulenko et~al\mbox{.}}{2019}]%
        {vakulenko2018qrfa}
\bibfield{author}{\bibinfo{person}{Svitlana Vakulenko}, \bibinfo{person}{Kate
  Revoredo}, \bibinfo{person}{Claudio Di~Ciccio}, {and}
  \bibinfo{person}{Maarten de Rijke}.} \bibinfo{year}{2019}\natexlab{}.
\newblock \showarticletitle{QRFA: A Data-Driven Model of Information-Seeking
  Dialogues}. In \bibinfo{booktitle}{\emph{ECIR}}.
\newblock


\bibitem[\protect\citeauthoryear{Van~Gysel, de~Rijke, and Kanoulas}{Van~Gysel
  et~al\mbox{.}}{2016}]%
        {van2016learning}
\bibfield{author}{\bibinfo{person}{Christophe Van~Gysel},
  \bibinfo{person}{Maarten de Rijke}, {and} \bibinfo{person}{Evangelos
  Kanoulas}.} \bibinfo{year}{2016}\natexlab{}.
\newblock \showarticletitle{Learning latent vector spaces for product search}.
  In \bibinfo{booktitle}{\emph{CIKM}}. \bibinfo{pages}{165--174}.
\newblock


\bibitem[\protect\citeauthoryear{Vaswani, Shazeer, Parmar, Uszkoreit, Jones,
  Gomez, Kaiser, and Polosukhin}{Vaswani et~al\mbox{.}}{2017}]%
        {vaswani2017attention}
\bibfield{author}{\bibinfo{person}{Ashish Vaswani}, \bibinfo{person}{Noam
  Shazeer}, \bibinfo{person}{Niki Parmar}, \bibinfo{person}{Jakob Uszkoreit},
  \bibinfo{person}{Llion Jones}, \bibinfo{person}{Aidan~N Gomez},
  \bibinfo{person}{{\L}ukasz Kaiser}, {and} \bibinfo{person}{Illia
  Polosukhin}.} \bibinfo{year}{2017}\natexlab{}.
\newblock \showarticletitle{Attention is all you need}. In
  \bibinfo{booktitle}{\emph{NeurIPS}}. \bibinfo{pages}{5998--6008}.
\newblock


\bibitem[\protect\citeauthoryear{Wan and McAuley}{Wan and McAuley}{2018}]%
        {wan2018item}
\bibfield{author}{\bibinfo{person}{Mengting Wan} {and} \bibinfo{person}{Julian
  McAuley}.} \bibinfo{year}{2018}\natexlab{}.
\newblock \showarticletitle{Item recommendation on monotonic behavior chains}.
  In \bibinfo{booktitle}{\emph{RecSys}}. \bibinfo{pages}{86--94}.
\newblock


\bibitem[\protect\citeauthoryear{Wang, Qiu, Huang, and He}{Wang
  et~al\mbox{.}}{2020a}]%
        {wang2020meta}
\bibfield{author}{\bibinfo{person}{Chengyu Wang}, \bibinfo{person}{Minghui
  Qiu}, \bibinfo{person}{Jun Huang}, {and} \bibinfo{person}{Xiaofeng He}.}
  \bibinfo{year}{2020}\natexlab{a}.
\newblock \bibinfo{title}{Meta Fine-Tuning Neural Language Models for
  Multi-Domain Text Mining}.
\newblock
\newblock
\showeprint[arxiv]{2003.13003}


\bibitem[\protect\citeauthoryear{Wang, Tang, Duan, Wei, Huang, Cao, Jiang,
  Zhou, et~al\mbox{.}}{Wang et~al\mbox{.}}{2020b}]%
        {wang2020k}
\bibfield{author}{\bibinfo{person}{Ruize Wang}, \bibinfo{person}{Duyu Tang},
  \bibinfo{person}{Nan Duan}, \bibinfo{person}{Zhongyu Wei},
  \bibinfo{person}{Xuanjing Huang}, \bibinfo{person}{Cuihong Cao},
  \bibinfo{person}{Daxin Jiang}, \bibinfo{person}{Ming Zhou}, {et~al\mbox{.}}}
  \bibinfo{year}{2020}\natexlab{b}.
\newblock \showarticletitle{K-adapter: Infusing knowledge into pre-trained
  models with adapters}.
\newblock \bibinfo{journal}{\emph{arXiv:2002.01808}} (\bibinfo{year}{2020}).
\newblock


\bibitem[\protect\citeauthoryear{Wu, Wu, Xing, Zhou, and Li}{Wu
  et~al\mbox{.}}{2017}]%
        {wu2017sequential}
\bibfield{author}{\bibinfo{person}{Yu Wu}, \bibinfo{person}{Wei Wu},
  \bibinfo{person}{Chen Xing}, \bibinfo{person}{Ming Zhou}, {and}
  \bibinfo{person}{Zhoujun Li}.} \bibinfo{year}{2017}\natexlab{}.
\newblock \showarticletitle{{Sequential Matching Network: A New Architecture
  for Multi-turn Response Selection in Retrieval-Based Chatbots}}. In
  \bibinfo{booktitle}{\emph{ACL}}, Vol.~\bibinfo{volume}{1}.
  \bibinfo{pages}{496--505}.
\newblock


\bibitem[\protect\citeauthoryear{Yang, Qiu, Qu, Guo, Zhang, Croft, Huang, and
  Chen}{Yang et~al\mbox{.}}{2018}]%
        {yang2018response}
\bibfield{author}{\bibinfo{person}{Liu Yang}, \bibinfo{person}{Minghui Qiu},
  \bibinfo{person}{Chen Qu}, \bibinfo{person}{Jiafeng Guo},
  \bibinfo{person}{Yongfeng Zhang}, \bibinfo{person}{W~Bruce Croft},
  \bibinfo{person}{Jun Huang}, {and} \bibinfo{person}{Haiqing Chen}.}
  \bibinfo{year}{2018}\natexlab{}.
\newblock \showarticletitle{{Response ranking with deep matching networks and
  external knowledge in information-seeking conversation systems}}. In
  \bibinfo{booktitle}{\emph{SIGIR}}. \bibinfo{pages}{245--254}.
\newblock


\bibitem[\protect\citeauthoryear{Yang, Xie, Lin, Li, Tan, Xiong, Li, and
  Lin}{Yang et~al\mbox{.}}{2019a}]%
        {yang2019end}
\bibfield{author}{\bibinfo{person}{Wei Yang}, \bibinfo{person}{Yuqing Xie},
  \bibinfo{person}{Aileen Lin}, \bibinfo{person}{Xingyu Li},
  \bibinfo{person}{Luchen Tan}, \bibinfo{person}{Kun Xiong},
  \bibinfo{person}{Ming Li}, {and} \bibinfo{person}{Jimmy Lin}.}
  \bibinfo{year}{2019}\natexlab{a}.
\newblock \showarticletitle{{End-to-End Open-Domain Question Answering with
  BERTserini}}. In \bibinfo{booktitle}{\emph{NAACL}}. \bibinfo{pages}{72--77}.
\newblock


\bibitem[\protect\citeauthoryear{Yang, Zhang, and Lin}{Yang
  et~al\mbox{.}}{2019b}]%
        {yang2019simple}
\bibfield{author}{\bibinfo{person}{Wei Yang}, \bibinfo{person}{Haotian Zhang},
  {and} \bibinfo{person}{Jimmy Lin}.} \bibinfo{year}{2019}\natexlab{b}.
\newblock \showarticletitle{{Simple applications of bert for ad hoc document
  retrieval}}.
\newblock \bibinfo{journal}{\emph{arXiv:1903.10972}} (\bibinfo{year}{2019}).
\newblock


\bibitem[\protect\citeauthoryear{Yuan, Zhou, Li, Lv, Zhu, Han, and Hu}{Yuan
  et~al\mbox{.}}{2019}]%
        {yuan2019multi}
\bibfield{author}{\bibinfo{person}{Chunyuan Yuan}, \bibinfo{person}{Wei Zhou},
  \bibinfo{person}{Mingming Li}, \bibinfo{person}{Shangwen Lv},
  \bibinfo{person}{Fuqing Zhu}, \bibinfo{person}{Jizhong Han}, {and}
  \bibinfo{person}{Songlin Hu}.} \bibinfo{year}{2019}\natexlab{}.
\newblock \showarticletitle{Multi-hop Selector Network for Multi-turn Response
  Selection in Retrieval-based Chatbots}. In
  \bibinfo{booktitle}{\emph{EMNLP-IJCNLP}}. \bibinfo{pages}{111--120}.
\newblock


\bibitem[\protect\citeauthoryear{Zamani and Croft}{Zamani and Croft}{2020}]%
        {zamani2020learning}
\bibfield{author}{\bibinfo{person}{Hamed Zamani} {and} \bibinfo{person}{W~Bruce
  Croft}.} \bibinfo{year}{2020}\natexlab{}.
\newblock \showarticletitle{Learning a Joint Search and Recommendation Model
  from User-Item Interactions}. In \bibinfo{booktitle}{\emph{CIKM}}.
  \bibinfo{pages}{717--725}.
\newblock


\bibitem[\protect\citeauthoryear{Zhang, Kishore, Wu, Weinberger, and
  Artzi}{Zhang et~al\mbox{.}}{2019}]%
        {zhang2019bertscore}
\bibfield{author}{\bibinfo{person}{Tianyi Zhang}, \bibinfo{person}{Varsha
  Kishore}, \bibinfo{person}{Felix Wu}, \bibinfo{person}{Kilian~Q Weinberger},
  {and} \bibinfo{person}{Yoav Artzi}.} \bibinfo{year}{2019}\natexlab{}.
\newblock \showarticletitle{BERTScore: Evaluating Text Generation with BERT}.
  In \bibinfo{booktitle}{\emph{ICLR}}.
\newblock


\bibitem[\protect\citeauthoryear{Zhang, Chen, Ai, Yang, and Croft}{Zhang
  et~al\mbox{.}}{2018}]%
        {zhang2018towards}
\bibfield{author}{\bibinfo{person}{Yongfeng Zhang}, \bibinfo{person}{Xu Chen},
  \bibinfo{person}{Qingyao Ai}, \bibinfo{person}{Liu Yang}, {and}
  \bibinfo{person}{W~Bruce Croft}.} \bibinfo{year}{2018}\natexlab{}.
\newblock \showarticletitle{Towards conversational search and recommendation:
  System ask, user respond}. In \bibinfo{booktitle}{\emph{CIKM}}.
  \bibinfo{pages}{177--186}.
\newblock


\bibitem[\protect\citeauthoryear{Zhang and Yang}{Zhang and Yang}{2017}]%
        {zhang2017survey}
\bibfield{author}{\bibinfo{person}{Yu Zhang} {and} \bibinfo{person}{Qiang
  Yang}.} \bibinfo{year}{2017}\natexlab{}.
\newblock \showarticletitle{A survey on multi-task learning}.
\newblock \bibinfo{journal}{\emph{arXiv:1707.08114}} (\bibinfo{year}{2017}).
\newblock


\bibitem[\protect\citeauthoryear{Zhou, Li, Dong, Liu, Chen, Zhao, Yu, and
  Wu}{Zhou et~al\mbox{.}}{2018}]%
        {zhou2018multi}
\bibfield{author}{\bibinfo{person}{Xiangyang Zhou}, \bibinfo{person}{Lu Li},
  \bibinfo{person}{Daxiang Dong}, \bibinfo{person}{Yi Liu},
  \bibinfo{person}{Ying Chen}, \bibinfo{person}{Wayne~Xin Zhao},
  \bibinfo{person}{Dianhai Yu}, {and} \bibinfo{person}{Hua Wu}.}
  \bibinfo{year}{2018}\natexlab{}.
\newblock \showarticletitle{Multi-turn response selection for chatbots with
  deep attention matching network}. In \bibinfo{booktitle}{\emph{ACL}}.
  \bibinfo{pages}{1118--1127}.
\newblock


\bibitem[\protect\citeauthoryear{Zhu, Kiros, Zemel, Salakhutdinov, Urtasun,
  Torralba, and Fidler}{Zhu et~al\mbox{.}}{2015}]%
        {zhu2015aligning}
\bibfield{author}{\bibinfo{person}{Yukun Zhu}, \bibinfo{person}{Ryan Kiros},
  \bibinfo{person}{Rich Zemel}, \bibinfo{person}{Ruslan Salakhutdinov},
  \bibinfo{person}{Raquel Urtasun}, \bibinfo{person}{Antonio Torralba}, {and}
  \bibinfo{person}{Sanja Fidler}.} \bibinfo{year}{2015}\natexlab{}.
\newblock \showarticletitle{Aligning books and movies: Towards story-like
  visual explanations by watching movies and reading books}. In
  \bibinfo{booktitle}{\emph{ICCV}}. \bibinfo{pages}{19--27}.
\newblock


\end{thebibliography}

\end{document}